\documentclass[useAMS,usenatbib,usegraphicx]{mn2e}
\pdfoutput=1
\usepackage{rotating}
\usepackage[3D]{movie15}
\usepackage{hyperref}
\usepackage{times}
\usepackage{caption}
\usepackage{lscape}
\usepackage{mathptmx}
\usepackage{amssymb}
\usepackage[fleqn]{amsmath}
\usepackage{url}
\usepackage{subfigure}
\usepackage{float}
\usepackage{enumitem}
\usepackage{txfonts}
\usepackage{xcolor}
\usepackage{bm}
\usepackage{natbib}
\usepackage{bibentry} 
\usepackage{etoolbox} 
\usepackage{graphicx}
\usepackage{epsfig}

\usepackage{xspace}

\newcommand{\Splash}{\textsc{splash}\xspace}
\newcommand{\Smoothviz}{\textsc{smoothviz}\xspace}
\newcommand{\MdotA}{\ifmmode{\dot{M}_{\eta_{\mathrm{A}}}}\else{$\dot{M}_{\eta_{\mathrm{A}}}$}\fi\xspace}
\newcommand{\MdotB}{\ifmmode{\dot{M}_{\eta_{\mathrm{B}}}}\else{$\dot{M}_{\eta_{\mathrm{B}}}$}\fi\xspace}
\newcommand{\etaA}{$\eta_{\mathrm{A}}$\xspace}
\newcommand{\etaB}{$\eta_{\mathrm{B}}$\xspace}
\newcommand{\ec}{$\eta$~Car\xspace}
\makeatletter
\newcommand\bibstyle@comma{\bibpunct{(}{)}{,}{a}{}{,}}
\newcommand\bibstyle@semicolon{\bibpunct{(}{)}{;}{a}{}{,}}
\makeatother

\pretocmd\citet{\citestyle{comma}}\relax\relax
\pretocmd\citep{\citestyle{semicolon}}\relax\relax

\newcommand{\Ms}{\ifmmode{~\mathrm{M}_{\odot}}\else{$\mathrm{M}_{\odot}$}\fi\xspace}
\newcommand{\Msy}{\ifmmode{\Ms \per{yr}}\else {$\Ms \per{yr}$}\fi\xspace}
\newcommand{\Ls}{\ifmmode{~\mathrm{L}_{\odot}}\else{$\mathrm{L}_{\odot}$}\fi\xspace}
\newcommand{\kms}{\ifmmode{~\mathrm{km}\per{s}}\else {$\mathrm{km}\per{s}$}\fi\xspace}
\newcommand{\Rs}{\ifmmode{~\mathrm{R}_{\odot}}\else{$\mathrm{R}_{\odot}$}\fi\xspace}
\newcommand{\per}[1]{\ifmmode{\mathrm{\,#1}^{-1}}\else {$\mathrm{\,#1}^{-1}$}\fi\xspace}

\newcommand{\ten}[1]{\ifmmode{10^{#1}}\else{$10^{#1}$}\fi\xspace}
\newcommand{\sci}[2]{\ifmmode{#1 \times 10^{#2}}\else{$#1 \times 10^{#2}$}\fi\xspace}

\newcommand{\HI}{\ifmmode{\mathrm{H\,I}}\else{H\textsc{$\,$i}}\fi\xspace}
\newcommand{\HII}{\ifmmode{\mathrm{H\,II}}\else{H\textsc{$\,$ii}}\fi\xspace}
\newcommand{\HeI}{\ifmmode{\mathrm{He\,I}}\else{He\textsc{$\,$i}}\fi\xspace}
\newcommand{\HeII}{\ifmmode{\mathrm{He\,II}}\else{He\textsc{$\,$ii}}\fi\xspace}
\newcommand{\HeIII}{\ifmmode{\mathrm{He\,III}}\else{He\textsc{$\,$iii}}\fi\xspace}
\newcommand{\FeIII}{\ifmmode{[\mathrm{Fe\,III}]}\else{[Fe\textsc{$\,$iii}]}\fi\xspace}
\newcommand{\FeII}{\ifmmode{[\mathrm{Fe\,II}]}\else{[Fe\textsc{$\,$ii}]}\fi\xspace}
\newcommand{\NiII}{\ifmmode{[\mathrm{Ni\,II}]}\else{[Ni\textsc{$\,$ii}]}\fi\xspace}
\newcommand{\nuHI}{\ifmmode{\nu_{\HI}}\else{$\nu_{\HI}$}\fi\xspace}
\newcommand{\nuHeI}{\ifmmode{\nu_{\HeI}}\else{$\nu_{\HeI}$}\fi\xspace}
\newcommand{\nuHeII}{\ifmmode{\nu_{\HeII}}\else{$\nu_{\HeII}$}\fi\xspace}

\defcitealias{madura13}{M13}
\defcitealias{clementel14}{C14}

\usepackage{color}

\title[3D printing $\eta$~Car's colliding winds]{3D printing meets computational astrophysics: deciphering the structure of $\eta$~Carinae's inner colliding winds}

\author[Madura et al.]{T.~I. Madura$^{1}$\thanks{NASA Postdoctoral Program Fellow}\thanks{E-mail: thomas.i.madura@nasa.gov}, N. Clementel$^{2}$, T.~R. Gull$^{1}$, C.~J.~H. Kruip$^{2}$ and J.-P. Paardekooper$^{3,4}$\\
$^{1}$Astrophysics Science Division, Code 667, NASA Goddard Space Flight Center, Greenbelt, MD 20771, USA\\
$^{2}$Leiden Observatory, Leiden University, PO Box 9513, 2300 RA Leiden, the Netherlands\\
$^{3}$Universit\"at Heidelberg, Zentrum f\"ur Astronomie, Institut f\"ur Theoretische Astrophysik, Alber--Ueberle--Str. 2, 69120 Heidelberg, Germany\\
$^{4}$Max Planck Institute for Extraterrestrial Physics, PO Box 1312, Giessenbachstr., D--85741 Garching, Germany\\
}

\begin{document}

\date{Accepted ***. Received ***; in original form ***}

\pagerange{\pageref{firstpage}--\pageref{lastpage}} \pubyear{2015}

\maketitle

\label{firstpage}

\begin{abstract}
We present the first 3D prints of output from a supercomputer simulation of a complex astrophysical system, the colliding stellar winds in the massive ($\gtrsim$120~\Ms), highly eccentric ($e \sim 0.9$) binary star system $\eta$~Carinae. We demonstrate the methodology used to incorporate 3D interactive figures into a PDF journal publication and the benefits of using 3D visualization and 3D printing as tools to analyze data from multidimensional numerical simulations. Using a consumer-grade 3D printer (MakerBot Replicator 2X), we successfully printed 3D smoothed particle hydrodynamics (SPH) simulations of $\eta$~Carinae's inner ($r \sim 110$~au) wind-wind collision interface at multiple orbital phases. The 3D prints and visualizations reveal important, previously unknown `finger-like' structures at orbital phases shortly after periastron ($\phi \sim 1.045$) that protrude radially outward from the spiral wind-wind collision region. We speculate that these fingers are related to instabilities (e.g. thin-shell, Rayleigh-Taylor) that arise at the interface between the radiatively-cooled layer of dense post-shock primary-star wind and the fast (3000~\kms), adiabatic post-shock companion-star wind. The success of our work and easy identification of previously unrecognized physical features highlight the important role 3D printing and interactive graphics can play in the visualization and understanding of complex 3D time-dependent numerical simulations of astrophysical phenomena.
\end{abstract}



\section{Introduction}\label{Intro}

The supermassive binary star system Eta~Carinae (\ec) is famous for the greatest non-terminal stellar explosion ever recorded \citep{davidson97}. In the 1840s, \ec became the second brightest non-solar-system object in the sky and ejected between 10 and 40~\Ms, forming the dusty bipolar ``Homunculus" nebula \citep{smith03, gomez10, steffen14}. Amazingly, this did not destroy the star(s). Multi-epoch ground- and space-based observations obtained over the past two decades reveal that \ec is a colliding wind binary (CWB) with a current total mass $\gtrsim$120~\Ms \citep{hillier01,hillier06} and a highly eccentric ($e \sim 0.9$), 5.54-yr orbit \citep{damineli97, whitelock04, damineli08a, damineli08b, gull09, groh10b, corcoran10, teodoro12}.

Because they are so luminous ($L_{\mathrm{Total}} \gtrsim \sci{5}{6} \Ls$, \citealt{hillier01, hillier06}), the stars in \ec have powerful radiation-driven stellar wind mass outflows. The Luminous Blue Variable (LBV) primary component, \etaA, has one of \emph{the} densest known stellar winds ($\MdotA \approx \sci{8.5}{-4} \Msy$, $v_{\infty} \approx 420 \kms$; \citealt{hillier01, groh12a}). The less luminous companion star, \etaB, has a much lower density, but faster, wind ($L_{\star} / \Ls \approx$~\ten{5}--\ten{6}, $\MdotB \approx \sci{1.4}{-5} \Msy$, $v_{\infty} \approx 3000 \kms$; \citealt{pittard02, parkin09}). These winds violently collide, producing strong X-ray emitting shocks \citep{pittard02, corcoran10, hamaguchi14} and a wind-wind interaction region (WWIR) that is thought to be the source of numerous forms of time-variable emission and absorption observed across a wide range of wavelengths \citep[see e.g.][]{damineli08b}.

Proper numerical modelling of \ec's WWIR remains a challenge, mainly because it requires a full three-dimensional (3D), time-dependent treatment since orbital motion, especially during periastron, greatly affects the geometry and dynamics of the WWIR. Three-dimensional hydrodynamical simulations of \ec show that the complex time-varying WWIR has a major impact on the observed X-ray emission \citep{okazaki08, parkin11, russell13}, the optical and ultraviolet (UV) light curves and spectra \citep{madura10, maduragroh12, madura12, madura13, clementel14, clementel15}, and the interpretation of various line profiles and interferometric observables \citep{groh10a, groh10b, groh12a, groh12b}. While 3D hydrodynamical simulations have helped to increase substantially our understanding of the present-day \ec binary, we are limited by an inability to adequately visualize the full 3D time-dependent geometry of the WWIR. However, this is crucial if we are to thoroughly understand how and where various forms of observed emission and absorption originate.

Most published figures of 3D simulations of \ec's colliding winds (and CWBs in general) consist of 2D slices through the origin of the typically-Cartesian 3D simulation domain, with color showing a scalar quantity such as density or temperature (see e.g. \citealt{lemaster07}, \citealt{okazaki08}, \citealt{parkin09}, \citealt{pittard09}, \citealt{parkin11}, and \citealt{madura13}, hereafter M13). Sometimes, 2D slices showing physical quantities in the two coordinate planes perpendicular to the orbital plane, in parallel planes above and below the orbital plane, and in planes at arbitrary angles relative to the orbital plane, are also provided (\citealt{lemaster07}, \citealt{okazaki08}, \citealt{pittard09}, and appendix~B of \citetalias{madura13}). While multi-panel figures showing 2D slices can be useful, the time-varying geometry of the WWIR (caused by orbital motion), combined with parameter studies of various stellar, wind, and orbital parameters, can lead to large numbers of cumbersome 2D figures. The amount of information that 2D slices can convey about an intrinsically 3D structure is also limited, which can make such 2D slices difficult to interpret and understand.

There have been attempts to provide 3D isovolume or isosurface renderings of the WWIR from 3D hydrodynamical simulations of CWBs, for example, fig.~3 of \citet{lemaster07} and figs.~2, 13, and 14 of \citet{pittard09}. For the specific case of \ec, a few 3D isovolume renderings exist in e.g. \citet{parkin11, maduragroh12}, and \citet{groh12b}. Such 3D renderings are typically the exception rather than the rule though, and, as is the case with 2D slices, multi-panel figures are necessary in order to show the full 3D structure of the simulation results from different viewing perspectives and/or at different orbital phases.

The predominance of 2D figures and animations in the literature is clearly driven by the need to display 3D data in a classic paper-journal format. In this sense, the problem of 3D visualization of complex simulations and observational data is not limited to \ec or CWBs. However, there is no real reason that researchers should be limited to 2D graphics when presenting their results in peer-reviewed publications. This is especially true since all major astrophysical journals are now published online. So-called `augmented articles' \citep{vogt13} are possible, in which 3D interactive models, images, sounds, and videos can be included directly within an \emph{Adobe Portable Document Format} (PDF) article. The inclusion of 3D interactive models in the astrophysics literature, via methods such as those described in \citet{barnes08}, are slowly becoming popular, and several astrophysical journals now fully support the inclusion of 3D interactive figures. Two very recent examples of such 3D interactive figures are fig.~1 of \citet{vogt14}, which presents a novel new way to classify galaxy emission lines via a 3D line ratio diagram, and fig.~5 of \citet{steffen14}, which presents a 3D interactive model of Eta~Carinae's bipolar `Homunculus' nebula that was constructed based on detailed spectral mapping observations obtained with the ESO Very Large Telescope/X-Shooter\footnote{For other recent examples, see references in \citet{vogt13}.}.

While the use of 3D interactive graphics will likely prove to be extremely helpful to astronomers in their quest for understanding and discovery, one should always try to keep an eye on emerging technologies that may further aid the astrophysical research community. One such technology that has been increasing in popularity across many different fields in recent years is additive manufacturing or `3D printing'. 3D printing has the potential to provide an entirely new method for researchers to visualize, understand, interpret, and communicate their science results.

The use of 3D printing in the astronomical community is still very much in its infancy, but several prominent examples have appeared within the past year. The first is a program aimed at transforming \emph{Hubble Space Telescope} (\emph{HST}) images into tactile 3D models using special software and 3D printers, with the goal of communicating the wonders of astronomy to the blind and visually-impaired \citep{christian14}. In the peer-reviewed literature there exists a 3D printable version of fig.~1 from \citet{vogt14} (see their fig.~15) and a 3D printable version of the Eta~Carinae Homunculus nebula model by \citet{steffen14}\footnote{See also \href{http://www.nasa.gov/content/goddard/astronomers-bring-the-third-dimension-to-a-doomed-stars-outburst/}{http://www.nasa.gov/content/goddard/astronomers-bring-the-third-dimension-to-a-doomed-stars-outburst/}}. Each of these is a unique illustration of how observational data can be used to develop a 3D printable model for increased understanding and communication of complex concepts. However, to date there have been no published attempts to use 3D visualization and printing techniques to aid in the understanding and communication of complicated multi-dimensional output from detailed numerical simulations of astrophysical phenomena.

In an effort to further demonstrate the benefits of using 3D visualization and 3D printing as tools to analyze output and communicate results from numerical simulations, we present the first 3D prints of output from a supercomputer simulation of a complex astrophysical system, the colliding stellar winds in the \ec binary. Using a consumer-grade 3D printer, we print output from 3D time-dependent smoothed particle hydrodynamics (SPH) simulations of \ec's inner ($r \sim 110$~au) wind-wind collision interface at multiple orbital phases. The main goal is to gain an improved understanding of the full 3D structure of \ec's WWIR and how it changes with orbital phase. The 3D prints and visualizations reveal previously unrecognized `finger-like' structures at orbital phases shortly after periastron that protrude radially outward from the spiral wind-wind collision region. The success of our work helps highlight the important role 3D printing can play in the visualization and understanding of complex 3D time-dependent simulations.

In the following section (\ref{sec:Method}), we describe our methodology and numerical approach, including the SPH simulations, the 3D visualization of the SPH output, and the generation of 3D printable files. Section~\ref{sec:Results} presents the results in the form of standard 2D images, pictures, and 3D interactive graphics. A brief discussion of the results and their implications is in Section~\ref{sec:Discussion}. Section~\ref{sec:Summary} summarizes our conclusions and outlines the direction of future work.


\section{Methods}\label{sec:Method}

\subsection{The 3D SPH simulations}\label{ssec:SPH}

The hydrodynamical simulation snapshots we visualize correspond to specific phases (apastron, periastron, and 3~months after periastron) from the Case~A ($\MdotA \approx \sci{8.5}{-4} \Msy$) and Case~C ($\MdotA \approx \sci{2.4}{-4} \Msy$) small-domain ($r = 10\,a = 155$~au) 3D SPH simulations of \citetalias{madura13}. In all simulations, the primary star's wind terminal speed is set to $420 \kms$, while the companion star mass loss rate and wind terminal speed are set to $\MdotB = \sci{1.4}{-5} \Msy$ and $v_{\infty} = 3000 \kms$, respectively. Additionally, the SPH particle mass is $5.913 \times 10^{24}$~g for the wind of \etaA and $5.913 \times 10^{23}$~g ($2.9565 \times 10^{24}$~g) for the wind of \etaB in the Case~A (Case~C) $r = 10a$ simulations. We refer the reader to \citetalias{madura13} and references therein for details on the SPH simulations and an extensive discussion of the results.

The spherical computational domain size of radius $r = 155$~au is chosen in order to investigate at sufficiently high spatial resolution the structure of \ec's inner WWIRs, since the `current' interaction between the two winds occurs at spatial scales comparable to the semi-major axis length $a \approx 15.4$~au. The three orbital phases selected are representative of when the WWIR has its simplest (apastron) and most complex (periastron and 3~months after periastron) geometry. The snapshot at apastron, when the stellar separation is largest and orbital speeds are lowest, provides a reference for the nearly-axisymmetric conical shape of the WWIR during most of the orbital cycle. The periastron snapshot defines the distorted geometry of the WWIR when the stellar separation is smallest and orbital speeds are greatest. The snapshot at 3~months after periastron ($\phi = 1.045$) corresponds to a time when the WWIR has a distinct Archimedean-spiral-like shape in the orbital plane due to the rapid orbital motion of the stars around periastron \citep{okazaki08, parkin11, madura12, madura13}. Simulations assuming two different \etaA mass loss rates are used to investigate how changes to the wind momentum ratio alter the WWIR opening angle, apex distance, 3D geometry, and dynamics.

We use a standard $xyz$ Cartesian coordinate system and set the orbit in the $xy$ plane, with the origin at the system centre of mass and the major axis along the $x$-axis. The stars orbit counter-clockwise when viewed from along the $+z$-axis. By convention, periastron is defined as $t = 0$ ($\phi = t/2024 = 0$). Simulations are started at apastron and run for multiple consecutive orbits.

\subsection{Grid construction and density distribution}\label{ssec:Grid}

SPH is a mesh-free method for solving the equations of fluid dynamics \citep{monaghan05} that is widely used in the astrophysical community. However, visualizing SPH data is far from straightforward since the data are highly adaptive and unstructured, defined on a set of points that follow the motion of the fluid. Simple interpolation to a uniform structured grid is often not an option since the grids are so immense in size they cannot be handled efficiently, or significant interpolation errors are introduced in areas of high particle density \citep{linsen11}. Due to these complications, several programs have been developed specifically for the visualization of SPH data. A popular, freely-available tool that allows for visualization of slices through the simulation volume, direct volume rendering, and particle rendering is \Splash \citep{price07}. Unfortunately, \Splash does not currently allow 3D isosurface extraction, which is required if we want to visualize \ec's stellar winds and WWIR as solid 3D surfaces. A more recent program designed for the interactive visual analysis of SPH data is \Smoothviz \citep{linsen11}. \Smoothviz allows isosurface extraction and direct volume rendering, but is currently limited to producing standard screenshots for use in research papers, as it uses OpenGL instead of graphics libraries such as \textsc{pgplot} \citep{molchanov13}.

Since our goal is to create 3D interactive figures and 3D printable files, we employ an alternative approach. We choose to generate a tetrahedral mesh from the SPH particle data in order to facilitate easier visualization with standard software such as \textsc{VisIt}\footnote{https://wci.llnl.gov/simulation/computer-codes/visit} and \textsc{ParaView}\footnote{http://www.paraview.org/}. The generation of tetrahedral meshes from particle data has a long tradition, with the widely accepted results of Delaunay tetrahedrization dating to 1934 \citep{du06, linsen11}. We therefore employ the same methodology as \citet{clementel14, clementel15} to generate our unstructured 3D mesh. Using the SPH particles themselves as the generating nuclei, we tessellate space according to the Voronoi recipe: all points in a grid cell are closer to the nucleus of that cell than to any other nucleus. The Voronoi nuclei are then connected by a Delaunay triangulation.

We assign to the nucleus of each Voronoi cell the corresponding SPH quantities of particle mass, density, temperature, and velocity, computed using the standard SPH cubic spline kernel \citep{monaghan92}. This helps ensure that each scalar variable visualized on our mesh closely matches that of the original SPH simulations, since the kernel samples a larger number of particles over a larger volume, resulting in quantities that are less affected by local differences in the SPH particle distribution \citep{clementel15}. Comparison with a direct visualization of the SPH density using \textsc{splash} \citep{price07} shows that this approach indeed matches well the density distribution of the original SPH simulations (see fig.~1 of \citealt{clementel15}). Fig.~2 of \citet{clementel15} shows an example of the unstructured mesh and number density at apastron for one of our 3D SPH simulations of \ec.

\subsection{Visualization}\label{ssec:Vis}

When visualizing scalar variables on our unstructured mesh, we would ideally like to render quantities that are centred on the original Voronoi cells that compose our 3D grid. Unfortunately, the Voronoi cells consist of a series of irregular $n$-sided polygons, which makes their visualization quite complex. Instead, it is much more straightforward to visualize the corresponding Delaunay triangulation. In 3D, the Delaunay cells are tetrahedra, which can be visualized using standard visualization tools. Since the Delaunay cells are tetrahedra, the quantity we visualize is the average of the four vertices that define the tetrahedron cell (i.e. the average of the four Voronoi nuclei). This approach works well for visualizing most physical quantities (e.g. temperature, density, velocity), and is suitable for our work. However, if neighbouring Voronoi nuclei have values which are significantly different (i.e. by several orders of magnitude), this `volume-average' approach may lead to tetrahedral-cell values that are difficult to understand and interpret (see \citealt{clementel15} for details).

To help the reader better appreciate the 3D structure of \ec's WWIR and the cavity carved within \etaA's wind by \etaB, we provide in Figs.~\ref{3dfig1}--\ref{3dfig6} three related visualizations (columns) of each SPH model. In each figure, the first column shows an arbitrary view looking down on the orbital plane, with the lower-density \etaB wind cavity opening toward (top row) or away from (bottom row) the observer. The top half of the 3D model ($z>0$) is transparent in order to clearly show the orbital plane. This view provides a useful reference for comparing the 3D results to 2D orbital-plane slices shown in earlier works such as \citetalias{madura13}.

Unfortunately, in a view such as that in the first column of Figs.~\ref{3dfig1}--\ref{3dfig6}, the fully rendered wind of \etaB prevents one from seeing the complete 3D geometry of the cavity carved within \etaA's wind. It would thus be useful to visualize the modified wind of \etaA separate from the lower-density \etaB wind. Doing this is straightforward since \etaA and \etaB have very different $\dot{M}$ (\MdotA / \MdotB$\approx 60$). Due to this $\dot{M}$ difference, our SPH simulations use different particle masses for each stellar wind. Therefore, using the SPH particle mass, we can isolate the pre- and post-shock \etaA winds and visualize them while keeping the entire pre- and post-shock \etaB winds, and the top ($z>0$) half of the model, transparent. Examples of this view are shown in the middle column of Figs.~\ref{3dfig1}--\ref{3dfig6}.

Since we are most interested in the 3D structure of \ec's WWIR, we must find a way to also visualize it separately from the individual stellar winds. For the WWIR, we choose to visualize the thin, dense post-shock \etaA wind region that is located between the contact discontinuity (which separates the colliding wind shocks) and the pre-shock \etaA wind. In order to isolate this specific region, we use its unique density and temperature distribution. Radiative cooling of the post-shock \etaA gas increases substantially its density, by at least an order of magnitude (\citealt{parkin11}; \citetalias{madura13}). There is thus a large difference in density between the pre- and post-shock \etaA winds. Since the SPH simulations use constant, spherical mass loss rates, the density dependence with radius from the stellar surface in each pre-shock stellar wind in the simulations is roughly given by $\rho_{\mathrm{spherical}}(r) = \dot{M} / [4 \pi r^{2} v(r)]$, where $v(r) = v_{\infty}(1 - R_{\star} / r)^{\beta}$ is the standard `beta-velocity law' ($\beta = 1$ for our simulations), with $v_{\infty}$ the wind terminal velocity and $R_{\star}$ the stellar radius \citepalias{madura13}. By defining a quantity $\delta \equiv \rho_{\mathrm{SPH}} / \rho_{\mathrm{spherical}}$, we can determine at each location the contrast in density between what is provided in the SPH simulations ($\rho_{\mathrm{SPH}}$), and what the density at that location would be in an undisturbed spherical stellar wind. Within the pre-shock wind, $\delta \approx 1$, while in the post-shock wind, $\delta$ is greater than one, usually much greater. Therefore, we isolate the dense, post-shock \etaA wind (from here on referred to generally as the WWIR) by computing $\delta$ for \etaA's wind and selecting only those regions with $\delta > 2$ and $T = 10,000$~K. Because the post-shock \etaA wind cools radiatively, it remains at the floor temperature $T = 10,000$~K set in the SPH simulations \citepalias{madura13}. Considering only regions with $T = 10,000$~K ensures that we isolate the WWIR from the much hotter ($T > \ten{6}$~K) post-shock \etaB gas. The last column of Figs.~\ref{3dfig1}--\ref{3dfig6} illustrates a view identical to that in the middle column of the figures, but with the addition of the 3D WWIR surface that exists above the orbital plane.

\subsection{Generation of 3D interactive figures}\label{ssec:3Dfigs}

So that readers can directly experiment with and see for themselves the full 3D structure of \ec's WWIR, we augment this article by incorporating directly into the PDF 3D interactive figures of both our SPH simulation results and the final 3D print models. We embed our 3D graphics directly into the PDF document in order to facilitate easy direct sharing of the 3D results. Nearly every figure in this paper has a 3D interactive counterpart that can be accessed using the freely-available software \emph{Adobe Acrobat Reader v.8.0}\footnote{http://get.adobe.com/reader} or above. Unfortunately, interactive 3D graphics in PDFs can currently only be viewed using \emph{Adobe Reader}. Other PDF viewers will display only the standard 2D images shown in each figure.

The 3D interactive graphics allow the reader to fully rotate, zoom, and fly around each 3D model. This is a very efficient tool for revealing the structure of \ec's WWIR. In some cases, the 3D figures allow for the display (or not) of different components of the model (e.g. Figs.~\ref{3dfig8}--\ref{3dfig10} and \ref{3dfig11}--\ref{3dfig13}), providing the reader control over what he/she wants to see. Pre-defined `views' to help guide the reader to specific orientations or features are also implemented, such as the orientation of the \ec binary on the sky as seen from Earth. We highly recommend after selecting and clicking on a specific 3D interactive figure, that the reader right-mouse-click the model and select from the available drop-down menu the option ``View in Floating Window". This will open the interactive 3D model in a small (although adjustable) side window that permits continued reading of the text and simultaneous viewing of other figures with minimal inconvenience. When in 3D interactive mode, right-click and select ``disable content'' to return to the original 2D figure. Numerous other options are available in the toolbar associated with each 3D figure and we encourage the reader to fully explore these.

To create our interactive 3D graphics, and in an effort to encourage others to use such 3D figures in their work, we rely on robust freely-available visualization software. The most difficult part is converting a particular 3D visualization into the U3D format required for embedding within the PDF document \citep{barnes08, vogt13}. We start by using either \textsc{VisIt} or \text{ParaView} to open and visualize our unstructured grid data. The choice of \textsc{VisIt} or \textsc{ParaView} is optional, and any suitable scientific visualization program may be used, provided it outputs the created data to the required format. Once we create our visualization, we export the model as either an OBJ or X3D file, depending on what we want to show. Both formats preserve the detailed geometry of the 3D models, but when converted to the U3D format, we find that X3D can preserve the color table used in a detailed scientific visualization, whereas OBJ sometimes does not. The downside to the X3D format is that the file size is usually larger than that of OBJ since the full detailed color information is being stored. The OBJ format is useful for situations where only the overall geometry, a solid surface color/transparency, and a small file size are needed.

Once an OBJ or X3D file is created it can be read directly into the freely-available 3D mesh processing software \textsc{MeshLab}\footnote{http://meshlab.sourceforge.net/}, where the 3D model can be adjusted and corrections applied if needed, and then exported directly to the required U3D format. Another option is to read the OBJ or X3D file into the professional (but still free) 3D rendering and animation software \textsc{Blender}\footnote{http://www.blender.org}. In \textsc{Blender}, textures and colors can be improved, added, subtracted, etc., and additional objects or meshes can be inserted (or removed). Numerous visualization possibilities exist with \textsc{Blender}, and we find that it is more stable and easier to use than \textsc{MeshLab}. However, \textsc{Blender} does not support exporting directly to the U3D format. Nonetheless, one can create their final 3D model for visualization using \textsc{Blender} and export it as either OBJ or X3D. Then, one need only use \textsc{MeshLab} to quickly and easily convert the OBJ/X3D file to the U3D format. In order to create the best possible visualizations, and in anticipation of creating 3D printable STL files (see Section~\ref{ssec:3Dprintfiles}), we employ \textsc{Blender} in this work, with a conversion to U3D using \textsc{MeshLab}. The above process for creating a U3D file may seem cumbersome, but we find that is in fact quite straightforward. The interested reader that does not require free software can alternatively purchase programs such as \textsc{PDF3DReportGen}\footnote{http://www.pdf3d.com/ } for the creation of U3Ds and 3D PDFs.

Once a U3D file is available, incorporating it into a PDF can be done using either standard commercial software or, as astronomers generally prefer, the free typesetting package \LaTeX{}. The \texttt{movie15} and \texttt{media9} \LaTeX{} style files fully support the embedding of 3D annotations in PDF documents. Using and calling the \texttt{movie15} and \texttt{media9} packages is incredibly straightforward and requires almost no more effort than inserting a standard 2D figure. For this paper, we use the \texttt{movie15} package.

\subsection{Generation of 3D print files and printing the results}\label{ssec:3Dprintfiles}

Having a 3D model and mesh, even one that looks nice on a computer screen, does not necessarily mean that it is suitable for 3D printing. Beside the requirement of converting the 3D model to the appropriate file format for use with a 3D printer (generally the STL format), a 3D design that is to be 3D printed must meet a few basic requirements. These ensure that the 3D model prints correctly.

A 3D design file to be printed must be closed, or ``watertight" as it is typically referred. All components in the 3D model should be connected to create a solid. Objects must typically also be manifold, i.e. have no edges that are shared between more than two faces. There should also be no parts of the model that have zero thickness. Individual and floating points, lines, and planes should be removed, since these do not have three separate directions necessary for printing. Surface normals also need to all be pointing in the same direction outward from the surface of the model. This ensures that the 3D printer does not `confuse' the internal and outer surfaces of the model. Finally, as may be obvious, the 3D model must have the appropriate physical dimensions to be printed on the specific printer of choice (i.e. your model must fit in the printer).

While different 3D printers and 3D `slicing' software may have more stringent requirements, we found in our work that the above were the minimal requirements necessary for a successful 3D print of our models. Creation of a 3D printable STL file follows nearly the same procedure as that outlined above for generating a U3D file for PDF display. One key distinction though is that most 3D printers are monochromatic, meaning they print using a single-color material per extruder. As such, the `color' of the 3D model to be printed does not matter, since typically only one or two solid colors at most will be available. Color information is also (usually) not stored in the STL file to be printed, and the choice of color is more of a manual `real-world' decision than something chosen in the file creation phase. Since we are mostly interested in the 3D geometry and dynamics of the WWIR, the main concern is preserving the overall 3D geometry of the design to be printed.

Thus, to create our 3D print files, we again visualize our SPH simulation data with \textsc{VisIt} or \textsc{ParaView}, but once satisfied with a particular visualization, we export our design in the OBJ file format. In order to fit our 3D model in our printer, while at the same time preserving as much interesting detail as possible, we crop the outer spherical edge and choose to visualize and print only the region extending to radius $r = 7a \approx 108$~au from the system centre-of-mass. This is a fairly small cropping of the model and, for the orbital phases of interest, no crucial information is lost.

Once we create an OBJ file, we import it into \textsc{Blender}, which has useful tools for detecting and correcting non-manifoldness, mesh holes, loose objects, and inverted normals. We find that in most cases, the most prominent problem with 3D OBJ files exported by \textsc{VisIt} or \textsc{ParaView} is that they have a large number of inverted normals. Luckily, these are easily corrected in \textsc{Blender} via a few simple mouse clicks. Once the normals are corrected, we use \textsc{Blender} to remove any floating points, lines, or faces that may lead to printing errors, and we close any open `holes' that make the model non-watertight. Here, we must be careful to mention that a large `hole' can exist in a model in a general physical sense and still allow that model to be 3D printable (see Results below). Instead, what matters is that the entire surface of the 3D model itself be closed, with all edges/faces connected.

To help guide the reader, using \textsc{Blender}, we add to our model two small spheres that are connected by a small thin cylinder, which represent the stars and illustrates their location and separation with respect to each other, the dense \etaA wind, and the WWIR. In order to make the individual stars more visible at the scale of our models, we have increased the radius of each sphere to be a factor 3.5 times larger than the correctly-scaled stellar radius. Thus, the spheres depicting the stars in Figs.~\ref{3dfig8}--\ref{fig14} have radii equivalent to 210 and 105~\Rs for \etaA and \etaB, respectively. The correct stellar separation to scale is used at each phase.

We make one other final adjustment in \textsc{Blender} before exporting our model to the STL format. This adjustment further ensures that our model fits the 3D printer and has a stable base on which to stand once it is printed. To provide an interesting scientific reference point for anyone viewing our 3D printed models, we rotate them to the correct derived orientation that the \ec binary has on the sky as seen from Earth, with an inclination $i = 138^{\circ}$, argument of periapsis $\omega = 263^{\circ}$, and position angle on the sky of the orbital angular momentum axis of PA$_{z} = 317^{\circ}$ (see \citealt{madura12}; \citetalias{madura13}). Once rotated, we remove a small portion of the bottom of the model so that it has a flat base. We are careful to not remove any of the WWIR or cavity carved within the \etaA wind by \etaB. Only a small portion of the undisturbed outer spherical \etaA wind is removed. This has no effect on our results or conclusions, but allows the 3D printed models to be placed on a flat surface and oriented to provide the viewer with an idea of how the system and WWIR appear on the sky, assuming North is up and East is left (see Fig.~\ref{fig14}).

To print our 3D models, we use a consumer-grade MakerBot Replicator 2X Experimental 3D Printer, which has dual-extrusion capabilities. We import our STL files into the freely-available \textsc{MakerWare}\footnote{https://www.makerbot.com/desktop} 3D printing software and create the X3G files specific to the MakerBot printer. Since our 3D models are incredibly complex and, in many cases, contain free-hanging unsupported edges, we print our models in one color using one of the MakerBot's extruders, and use the second extruder to print dissolvable support material. Once printed, we place our model in a limonene bath, which safely dissolves away the support material. We use the highest layer resolution possible when printing (100~microns), and a physical size that occupies nearly the entire available build volume (model diameter $\approx 6$~inches $\approx 15$~cm across its widest part).

Each 3D printed model consists of two parts, joined by small metal pins. The bottom half of each model consists of the dense \etaA wind and the hollow \etaB wind cavity, while the top half consists solely of the WWIR (dense post-shock \etaA wind region). The two pieces are separable to allow the viewer to see the internal regions of the cavity carved by \etaB in \etaA's wind, or the WWIR by itself (see e.g. Figs.~\ref{3dfig10} and \ref{3dfig13}). Additionally, we add two small beads connected by a short pin to represent the individual stars and illustrate their orientation and separation. The radii of the beads and their separation is to scale with the rest of the printed model, although as described above, the stellar radii have been increased by a factor of 3.5 to make the stars more easily visible.

Finally, we note that each 3D printable STL file is attached to this article as supplementary online material. Thus, anyone with access to a suitable 3D printer can in principle print their own \ec wind model(s). The STL format should be compatible with most, if not all, 3D printers currently on the market. The STL file should be `ready to print' without additional modifications, but due to differences in printers and printing software, we cannot guarantee that absolutely no modifications are necessary. At the very least, the model will need to be adjusted to physically fit the 3D printer being used. We also have only tried printing the models on a 3D printer with dual extrusion and dissolvable support material, and are unsure of the results of using a single-extrusion printer. We welcome readers with questions about 3D printing the models, the model creation/design, and/or the file creation process to contact us directly.


\section{Results}\label{sec:Results}

\begin{figure*}
\centering \includemovie[
     3Dviews2=views_JoinedMesh.tex,
        toolbar, 
        label=Fig1.u3d,
     text={\includegraphics[width=174mm]{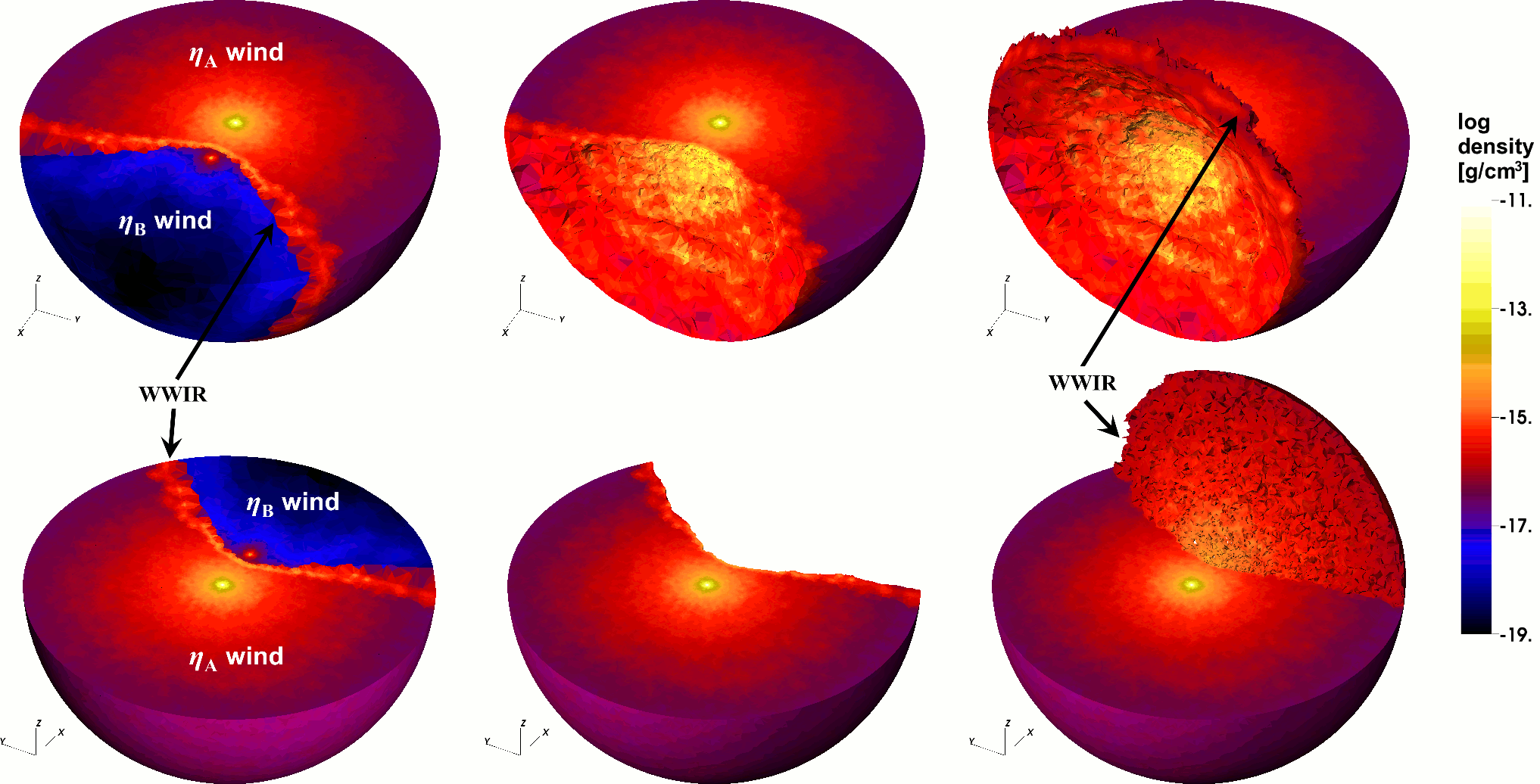}},
        3Dlights=CAD,
]{}{}{Fig1.u3d}
\caption{3D view of \ec's WWIR at apastron for the Case~A ($\MdotA = \sci{8.5}{-4} \Msy$) SPH simulation. The top left panel shows an (arbitrary) view looking down on the orbital plane, with the lower-density \etaB wind cavity opening toward the observer. The top half of the 3D model is transparent in order to clearly show the orbital plane. The top middle panel is the same as the first, but with the lower-density \etaB wind made completely transparent in order to better show the 3D structure of the \etaB wind cavity. The top right panel is identical to the top middle panel, but also includes the 3D surface of the WWIR that exists above the orbital plane. Panels in the bottom row are identical to those in the top row, but show a 180$^{\circ}$ rotated view. Color shows log density in cgs units in all panels. The locations of the stellar winds and WWIR are indicated. Click the image for a 3D interactive version of the model view shown in the last column (Adobe Reader$^{\circledR}$ only. We suggest selecting ``View in Floating Window" in the right-mouse-click drop-down menu). When in 3D interactive mode, right-click and select ``disable content'' to return to the original 2D figure. A small white sphere in the 3D interactive model marks the location of the companion star \etaB.}
\label{3dfig1}
\end{figure*}

\begin{figure*}
\centering \includemovie[
    3Dviews2=views_JoinedMesh.tex,
        toolbar, 
        label=Fig2.u3d,
     text={\includegraphics[width=174mm]{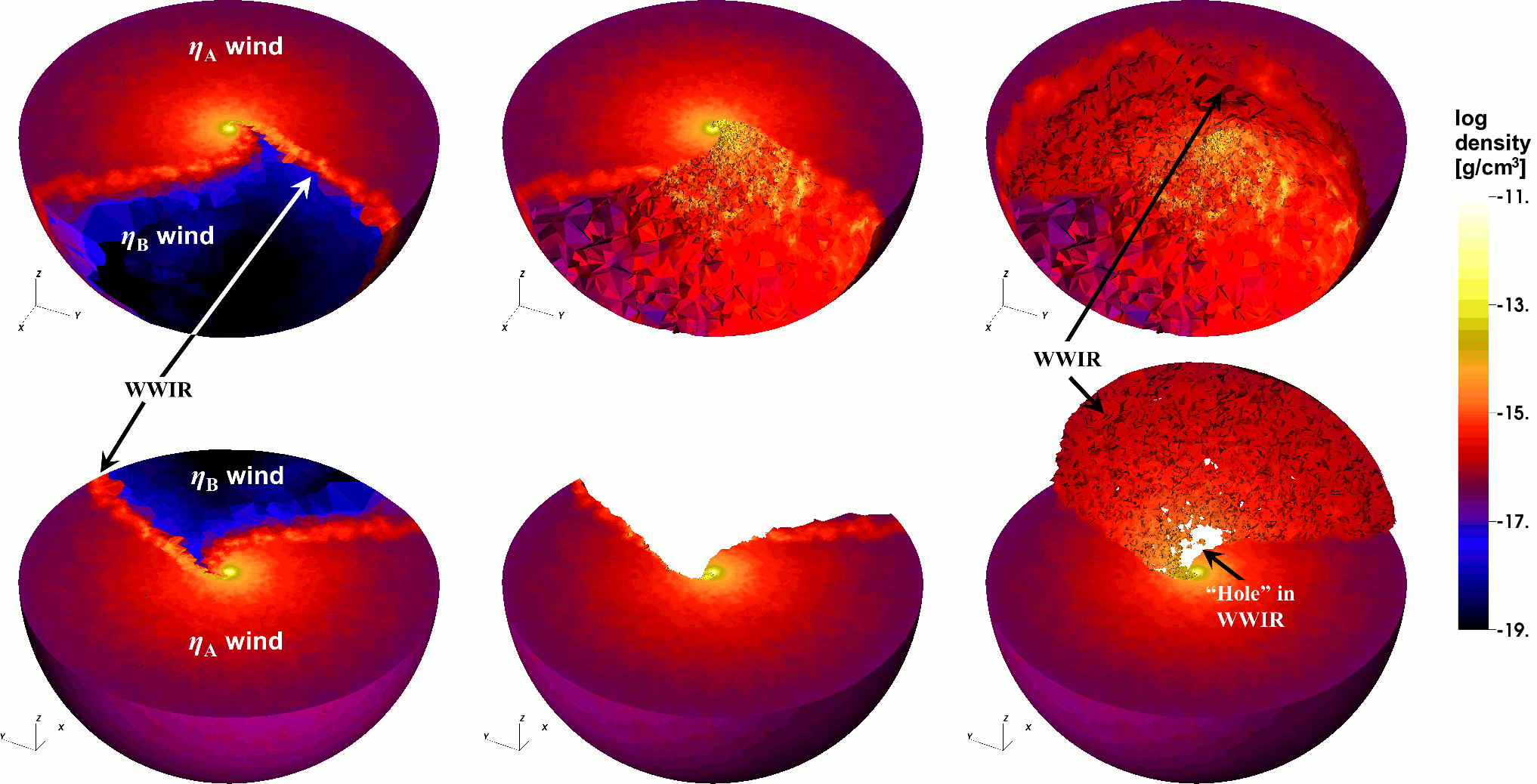}},
        3Dlights=CAD,
]{}{}{Fig2.u3d}
\caption{Same as Fig.~\ref{3dfig1}, but at periastron. Click image for a 3D interactive model (Adobe Reader$^{\circledR}$ only).}
\label{3dfig2}
\end{figure*}

\begin{figure*}
\centering \includemovie[
     3Dviews2=views_JoinedMesh.tex,
        toolbar, 
        label=Fig3.u3d,
     text={\includegraphics[width=174mm]{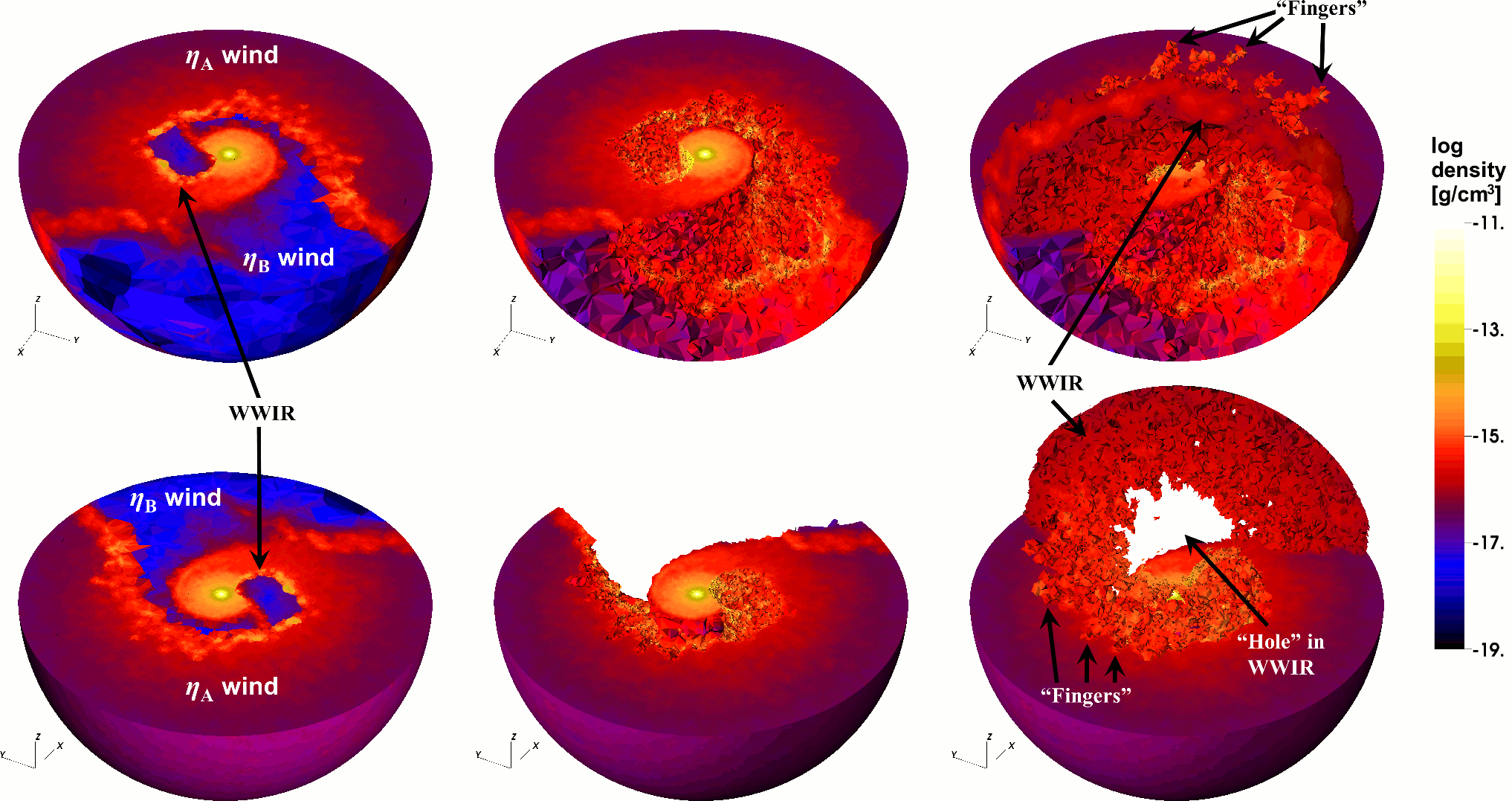}},
        3Dlights=CAD,
]{}{}{Fig3.u3d}
\caption{Same as Fig.~\ref{3dfig1}, but at $\approx$~3~months after periastron ($\phi = 1.045$). Click image for a 3D interactive model (Adobe Reader$^{\circledR}$ only).}
\label{3dfig3}
\end{figure*}

\begin{figure*}
\centering \includemovie[
     3Dviews2=views_JoinedMesh3.tex,
        toolbar, 
        label=Fig4.u3d,
     text={\includegraphics[width=174mm]{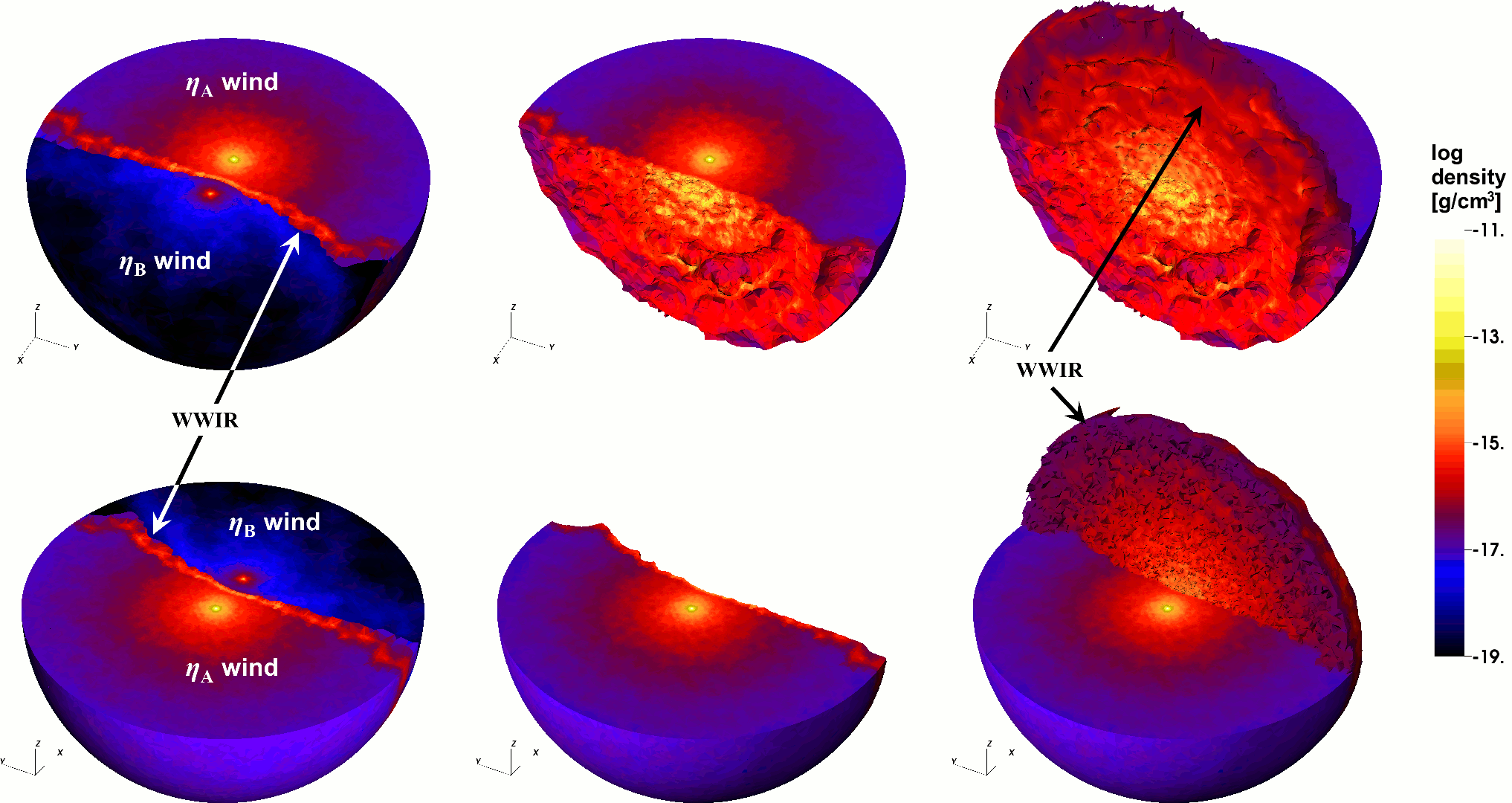}},
        3Dlights=CAD,
]{}{}{Fig4.u3d}
\caption{Same as Fig.~\ref{3dfig1}, but for the Case~C SPH simulation ($\MdotA = \sci{2.4}{-4} \Msy$). Click image for a 3D interactive model (Adobe Reader$^{\circledR}$ only).}
\label{3dfig4}
\end{figure*}

\begin{figure*}
\centering \includemovie[
     3Dviews2=views_JoinedMesh.tex,
        toolbar, 
        label=Fig5.u3d,
     text={\includegraphics[width=174mm]{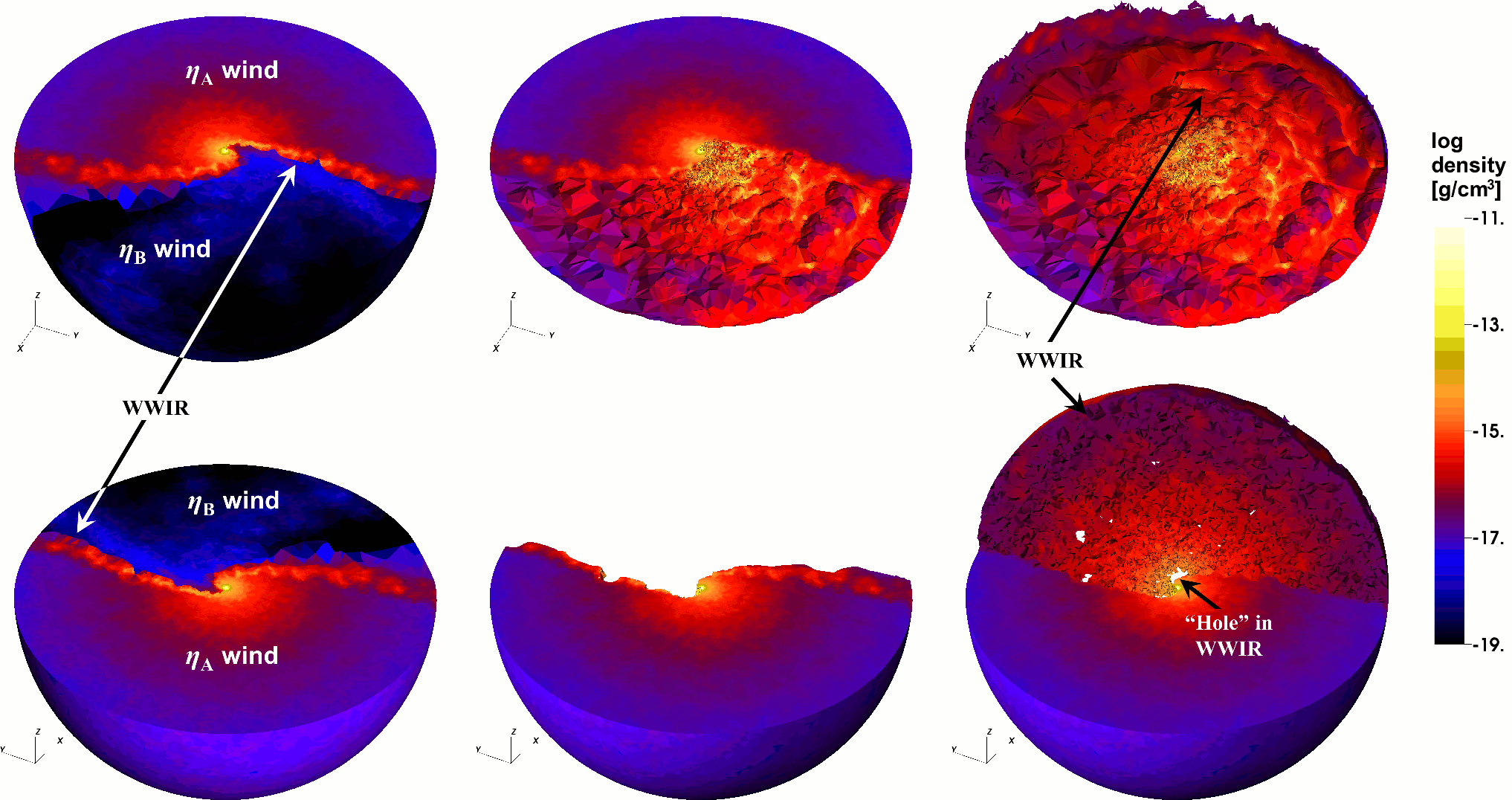}},
        3Dlights=CAD,
]{}{}{Fig5.u3d}
\caption{Same as Fig.~\ref{3dfig4}, but at periastron. Click image for a 3D interactive model (Adobe Reader$^{\circledR}$ only).}
\label{3dfig5}
\end{figure*}

\begin{figure*}
\centering \includemovie[
     3Dviews2=views_JoinedMesh.tex,
        toolbar, 
        label=Fig6.u3d,
     text={\includegraphics[width=174mm]{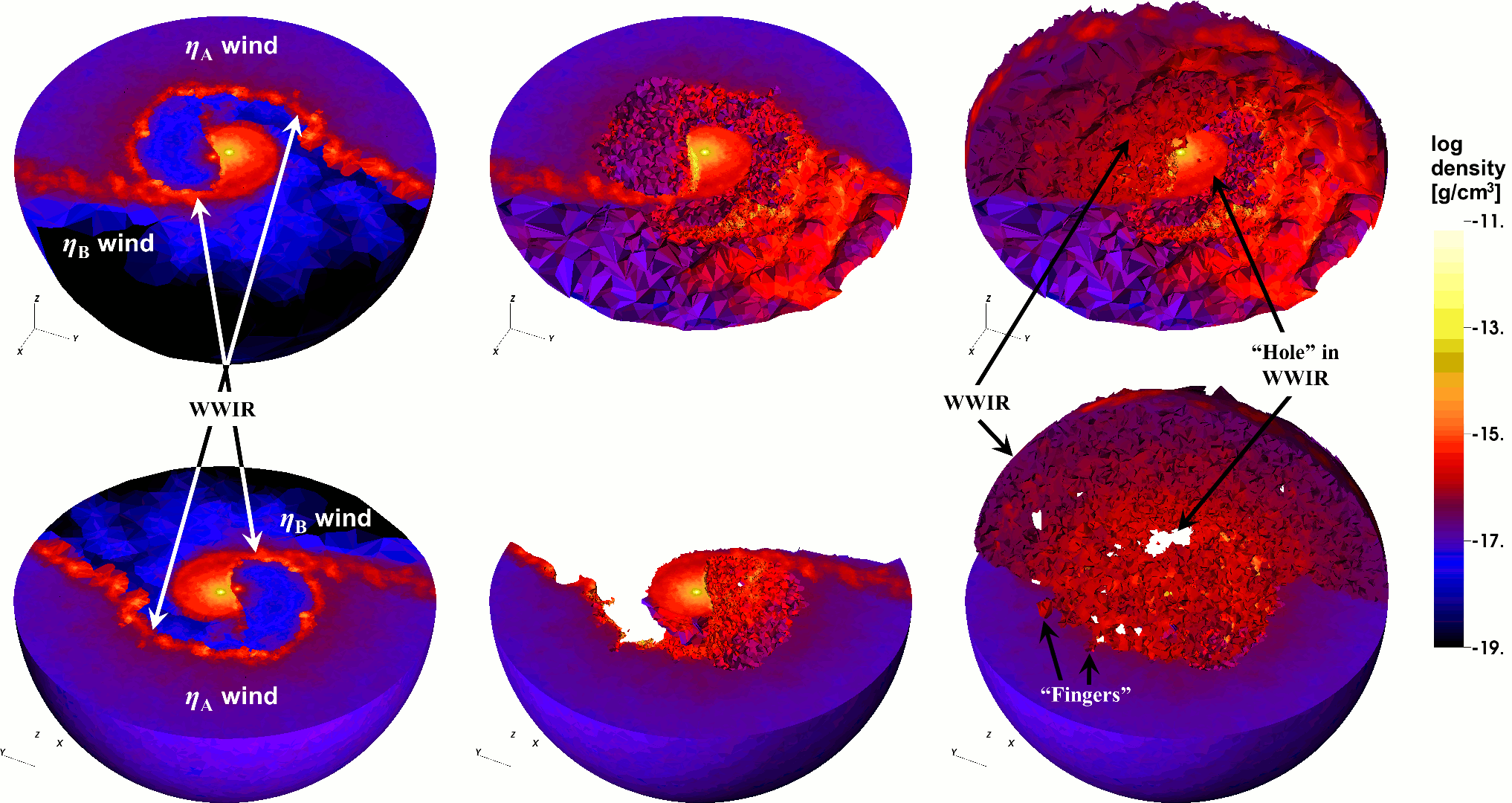}},
        3Dlights=CAD,
]{}{}{Fig6.u3d}
\caption{Same as Fig.~\ref{3dfig4}, but at $\approx$~3~months after periastron ($\phi = 1.045$). Click image for a 3D interactive model (Adobe Reader$^{\circledR}$ only).}
\label{3dfig6}
\end{figure*}

\begin{figure*}
\includemovie[
 controls=true,
text={\includegraphics[width=160mm]{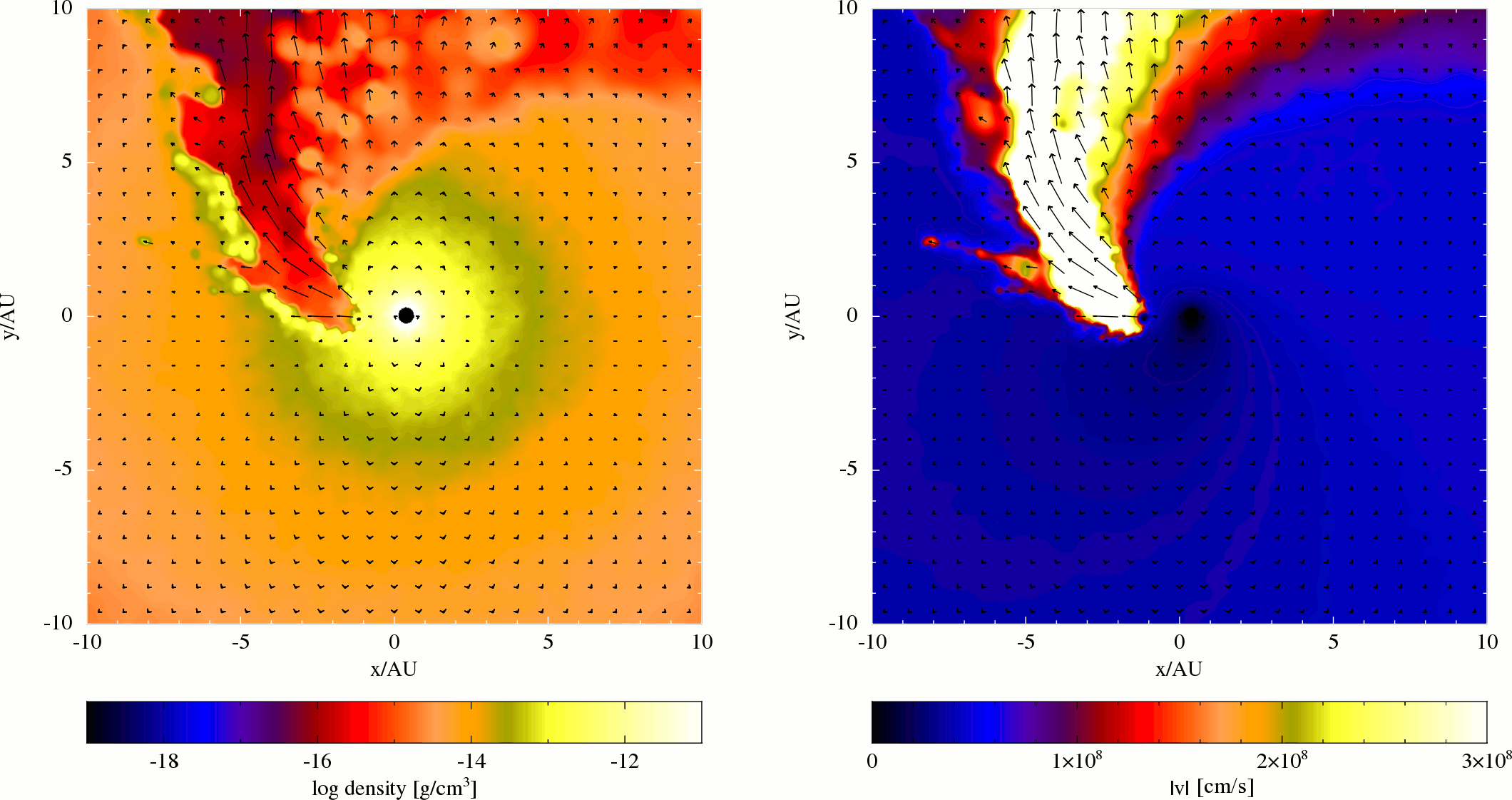}}]
 {\linewidth}{0.5\linewidth}{Fig7Movie.mp4}
\caption{Density (left) and wind speed (right) in the orbital plane at periastron for the small-domain ($r = 1.5a$) Case~A simulation of \citetalias{madura13}. Wind velocity vectors (arrows) are overlaid on both plots. The length of the arrows is proportional to the magnitude of the wind speed. Click the figure to play a short movie showing the evolution of the density and wind velocity in the orbital plane. The movie starts at orbital phase 0.95 ($\sim 100$ days before periastron) and ends at phase 1.05 ($\sim100$ days after periastron). The movie frame rate is set to 15 frames/second in order to better show the evolution of the wind velocity.}
 \label{fig7}
\end{figure*}

\subsection{The 3D visualizations}\label{ssec:3Dfigres}

Figs.~\ref{3dfig1}--\ref{3dfig6} present the results of our 3D visualizations of \ec's inner WWIR. Embedded within each 2D figure is an interactive 3D version of the model view shown in the last column. Figs.~\ref{3dfig1}--\ref{3dfig3} correspond to the higher \MdotA Case~A simulation, while Figs.~\ref{3dfig4}--\ref{3dfig6} correspond to the lower \MdotA Case~C simulation. Case~A is more representative of \etaA's current $\dot{M}$ \citepalias{madura13}, while Case~C is included mainly to investigate how changes to \MdotA affect the overall 3D geometry of the WWIR. Each model shown has a radius $r = 7a \approx 108$~au, measured from the system centre-of-mass.

Figures~\ref{3dfig1} and \ref{3dfig4} show that at apastron, the WWIR and cavity carved by \etaB in \etaA's dense wind have the expected nearly-axisymmetric conical shape, with the WWIR opening angle increasing with decreasing \MdotA \citepalias{madura13}. However, one does not fully appreciate how dramatic the change in WWIR opening angle is until it is visualized in 3D. In Fig.~\ref{3dfig1}, the WWIR is conical, whereas in Fig.~\ref{3dfig4}, it is nearly a plane. Another interesting feature more clearly visible in the 3D renderings is the very slight distortion of the WWIR in the direction of orbital motion. This is due to the additional component to the wind velocities caused by the non-zero velocity of the stars about the system centre-of-mass.

We also see in Figs.~\ref{3dfig1} and \ref{3dfig4} that the WWIR is not a clean, smooth surface. Rather, it is corrugated and contains many smaller-scale bumps and protrusions that arise as a result of various instabilities, such as non-linear thin shell and Kelvin-Helmholtz (\citealt{parkin11}; \citetalias{madura13}). While somewhat difficult to see in the density-colored models of Figs.~\ref{3dfig1} and \ref{3dfig4}, Figs.~\ref{3dfig8} and \ref{3dfig11} clearly show that the WWIR in the Case~C simulation appears to contain much more detailed structure, including large trenches that extend into and around the surface of the WWIR that faces the observer. We are unsure if this means that the Case~C result is intrinsically more unstable, or if the WWIR is simply better resolved as this SPH simulation used more SPH particles (by a factor of $\sim 1.5$) and had a slightly better numerical resolution.

Figures~\ref{3dfig2} and \ref{3dfig5} illustrate the twisted 3D geometry of the WWIR at periastron. In both figures we find that the leading arm of the WWIR (near the apex) is highly twisted in the direction of orbital motion, whereas the spatially-extended outer remnant of the WWIR trailing arm, created before periastron, maintains some of its initial axisymmetric geometry. There is also more curvature of the WWIR apex in the Case~A simulation than in the Case~C simulation. In Case~C (fig.~\ref{3dfig5}), the WWIR is still amazingly flat and planar in shape, although with a `twist' of the apex and a spatial displacement of the leading edge of the WWIR further into the pre-shock \etaA wind.

One particularly interesting feature in the WWIR at periastron, visible in both \MdotA cases, is the hole in the trailing arm near the WWIR apex. This WWIR hole is prominently located above (and below) the orbital plane, and is something not noticed in previous works that relied on 2D image slices through the orbital plane. This hole is a direct result of the fact that, at those locations where the hole exists, there is no longer a WWIR. We emphasize that it is not due to our choice of parameters used to isolate the WWIR. Plots of the temperature show that there is no shock-heated gas in regions where the hole exists. The temperature of material in the WWIR hole is $\approx 10^{4}$~K. Plots of the $\delta$ parameter also show no density enhancement for material where the WWIR hole is located. Material within the WWIR hole has $\delta \lesssim 1$.

Instead, because of the extremely high orbital eccentricity and embedding of \etaB within \etaA's dense wind at periastron, \etaB's outwardly expanding wind is unable to collide with \etaA's downstream wind. The wind of \etaB becomes trapped in specific directions at periastron and there is no longer any wind-wind collision at certain locations in the trailing wind. As a result, the unshocked primary wind starts expanding and filling in the low-density cavity that was carved by \etaB's wind during the broad part of the orbit. The outermost WWIR-trailing-arm located downstream from the hole remains intact at periastron simply because of time-delay effects; the hole near the apex has not had enough time to expand and propagate downstream.

The physical reasons for the formation of the WWIR hole are demonstrated in Fig.~\ref{fig7}, which shows the density and wind speed in the orbital plane at periastron for the small-domain ($r = 1.5a$) Case~A simulation of \citetalias{madura13}, with the wind velocity vectors (arrows) overlaid. The length of the arrows is proportional to the magnitude of the wind speed. If the reader clicks the figure, a short movie plays showing the evolution of the density and wind velocity in the orbital plane, starting at orbital phase 0.95 ($\sim 100$ days before periastron), continuing through periastron and ending at phase 1.05 ($\sim100$ days after periastron).

The animation shows that leading up to periastron, there is a typical wind-wind collision. However, shortly before and at periastron, the stellar separation is small enough that, combined with radiative inhibition effects, \etaB can no longer effectively drive a wind toward \etaA \citepalias{madura13}. The dense wind of \etaA thus overwhelms \etaB, which can now only drive a stellar wind in directions away from \etaA. Fig.~\ref{fig7} shows that at periastron, all of the wind vectors point away from \etaA, and \etaB is deeply embedded within \etaA's extremely dense inner wind. Thus, there can no longer be a wind-wind collision in the trailing arm downstream from the stars (downstream in the direction opposite that of the orbital motion).

The results of Fig.~\ref{fig7} demonstrate that the WWIR hole is not due to an inadequate numerical resolution of the simulations, since Fig.~\ref{fig7} uses the highest-resolution, smaller-domain simulation from \citetalias{madura13}. This simulation uses 8 times the number of particles as the $r = 10a$ simulations shown in the figures throughout the rest of this paper. The SPH particle mass is $5.913 \times 10^{23}$~g for the wind of \etaA and $1.971 \times 10^{23}$~g for the wind of \etaB in the $r=1.5a$ simulations. Although different SPH particle masses are used for the individual stellar winds since \etaA's mass loss rate is much higher than that of \etaB, the $r = 1.5a$ simulations are of adequate resolution to resolve both stellar winds, as demonstrated in \citetalias{madura13}.

We find that the appearance and size of the WWIR hole does depend on the mass loss rate used for \etaA. The hole at periastron is larger and more prominent in Case~A than Case~C. As \etaA's mass loss rate is lowered, the hole forms later, closer to periastron, and is smaller in overall size after periastron. This is consistent with the change in wind momentum ratio and WWIR opening angle as \MdotA is lowered. The fast wind from \etaB is able to better compete against \etaA's wind in Case~C, and thus able to better maintain the trailing arm of the WWIR at periastron. Lower \MdotA also delay and alter any WWIR collapse that may occur around periastron \citepalias{madura13}. This is consistent with the simulation results of \citet{parkin11}, since their simulations assumed a primary star mass loss rate that is roughly half that which we use for the simulations in Figs.~\ref{3dfig1}--\ref{3dfig3}, \ref{fig7}, and \ref{3dfig8}--\ref{3dfig10}, in which the WWIR hole is most prominent.

Interestingly, there is no evidence of a WWIR hole around periastron in the 3D \ec simulations of \citet{parkin11}. There are several reasons for this, discussed extensively in Section~3.1.1 and Appendix~A2 of \citetalias{madura13} (to which we refer the reader for details). The key point is that the stellar radius and wind-velocity law assumed for \etaB, together with radiative cooling effects, have a big influence on determining whether there will be a cooling-transition phase from adiabatic to radiative in \etaB's post-shock wind around periastron. The slightly larger stellar radius and beta-wind velocity parameter used in our simulations lead to a significant reduction in the pre-shock \etaB wind speed around periastron, which results in strong, rapid cooling of the post-shock \etaB gas. Using the \etaB parameters of \citet{parkin11}, the reduction in \etaB's wind speed by \etaA is insufficient to cause \etaB's post-shock wind to switch strongly to the radiative-cooling regime and cause a WWIR collapse. \etaB's wind in the simulations of \citet{parkin11} is thus able to effectively maintain a wind-wind collision during periastron passage, even downstream, explaining why no hole is seen in the trailing arm of the WWIR in their simulations. Therefore, the existence of the WWIR hole around periastron depends strongly on the assumed stellar, wind, and orbital parameters of the \ec system.

Of the three phases studied, that at $\sim 3$~months after periastron ($\phi = 1.045$, Figs.~\ref{3dfig3}, \ref{3dfig6}, \ref{3dfig10}, and \ref{3dfig13}) contains the most fascinating 3D WWIR geometry. As expected based on previous 3D simulations, a spiral cavity is carved within the dense wind of \etaA by \etaB's fast wind. However, the full 3D geometry of this cavity has never been visualized before, and we observe some interesting new features and phenomena that occur in both \MdotA simulations.

First, as alluded to above, as the 3D geometry of the WWIR evolves with time following periastron, the hole created near the WWIR apex at periastron expands and propagates downstream along the WWIR's old trailing arm. Thus, the new spiral WWIR created during periastron essentially has no trailing arm, and consists predominately of a leading arm and structures above and below the orbital plane. The hole created at periastron grows as the system moves through periastron and slowly eats its way into the remnant of the WWIR that was created during the broad part of the orbit. This effect is more prominent the higher the value of \MdotA.

Next, we see that the spiral wind cavity carved by \etaB within the back side of \etaA's wind during periastron passage is much shallower than the cavity carved during the broad part of the orbit. This is due to the short amount of time \etaB spends on the far side of \etaA during periastron. The spiral wind cavity and new WWIR formed during periastron have yet to expand outward to a size comparable to the remnant cavity on the apastron side of the system.

The most surprising new set of features found at $\phi = 1.045$ are the protrusions or `fingers' that extend radially from the spiral WWIR (see Figs.~\ref{3dfig3}, \ref{3dfig6}, \ref{3dfig10}, and \ref{3dfig13}). Detailed examination reveals that these fingers are actually tubes that consist of a thin shell of cold ($T \sim 10,000$~K), dense, compressed post-shock \etaA wind that is filled with hotter ($T \gtrsim 10^5$~K) post-shock \etaB wind. The fingers penetrate into the unshocked \etaA wind expanding on the periastron side of the system and extend slightly above and below the orbital plane from the leading arm of the WWIR. They do not extend perfectly vertically above or below the orbital plane (i.e. they are not at an angle of $90^{\circ}$ with respect to the orbital plane). Instead, they point radially outward away from the stars. The fingers extend in the same direction that \etaB's fast wind is able to collide with \etaA's wind during periastron passage, and their location appears to be tied to the direction and speed of orbital motion around periastron.

Analysis of the high-resolution $r = 1.5a$ simulation shows that the first fingers start to develop at orbital phase $\phi \approx 0.993$ (about 2 weeks before periastron). Additional fingers continue to develop until $\phi \approx 1.01$ (about 3 weeks after periastron). Approximately two dozen noticeable fingers are present on the entire 3D surface of a spiral WWIR once periastron passage is complete (i.e. at $\phi \gtrsim 1.01$). The total number and distribution of fingers varies from spiral to spiral due to the nature of the instabilities that form them. The spacing of the fingers is generally comparable to a few times the thickness of the dense WWIR shell, although in some locations the spacing is approximately equal to the thickness of the WWIR shell.

The length of the fingers is also impressive, being larger than the stellar separation at this phase (i.e. $\gtrsim 7$~au). Their diameter is on the order of several au. The fingers never completely vanish in our simulations, but slowly expand and cool, increasing in volume and moving outward until they eventually leave the computational domain. The expanded fingers are noticeable in the outermost shells of compressed \etaA wind located to the left in the large-scale SPH simulations shown in figures~8, B5, and B7 of \citetalias{madura13}. These outer expanded fingers are quite large, in some cases larger than the entire central binary orbit. The gas within them has also cooled adiabatically down to the floor temperature of the simulations (10,000~K).

We also find that \MdotA has some interesting effects on the 3D geometry of the WWIR at $\phi =$1.045. The WWIR in Case~C appears to have fewer intact fingers. The protrusions present in Case~C also appear to be shorter compared to those in Case~A. Curiously, while the Case~C WWIR lacks protrusions, it has an abundance of holes. We suspect that these holes in the WWIR are actually indicative of protrusions, ones in which the dense outer shell of post-shock \etaA wind that defines the surface of the protrusion/tube has a $\delta < 2$. This idea is supported by examination of the wind cavity carved within \etaA's dense wind (Fig.~\ref{3dfig12}), which exhibits small well-defined tubular cavities that extend into the outwardly expanding unshocked \etaA wind. Moreover, the holes in the Case~C WWIR all line up and point in the same direction as the protrusions that are present.

The spiral WWIR and wind cavity in Case~C are also noticeably broader than in Case~A (Fig.~\ref{3dfig13}). This is not surprising and is a result of the larger opening angle due to the lower \MdotA in Case~C. What is intriguing is that in Case~C, the outermost part of the leading arm of the `current' WWIR between/around the stars is so large that it has caught up to and started to collide and merge with the remnant of the old trailing arm that formed long before periastron. In Case~A at $\phi =$1.045, the current WWIR has yet to reach the remains of the old trailing arm and there is still a large separation between it and the remnants of the trailing arm.

\subsection{3D printing results}\label{ssec:3Dprintres}

\begin{figure}
\centering \includemovie[
     3Dviews2=views_ApastronHighMdot.tex,
        toolbar, 
        label=Fig8.u3d,
     text={\includegraphics[width=83mm]{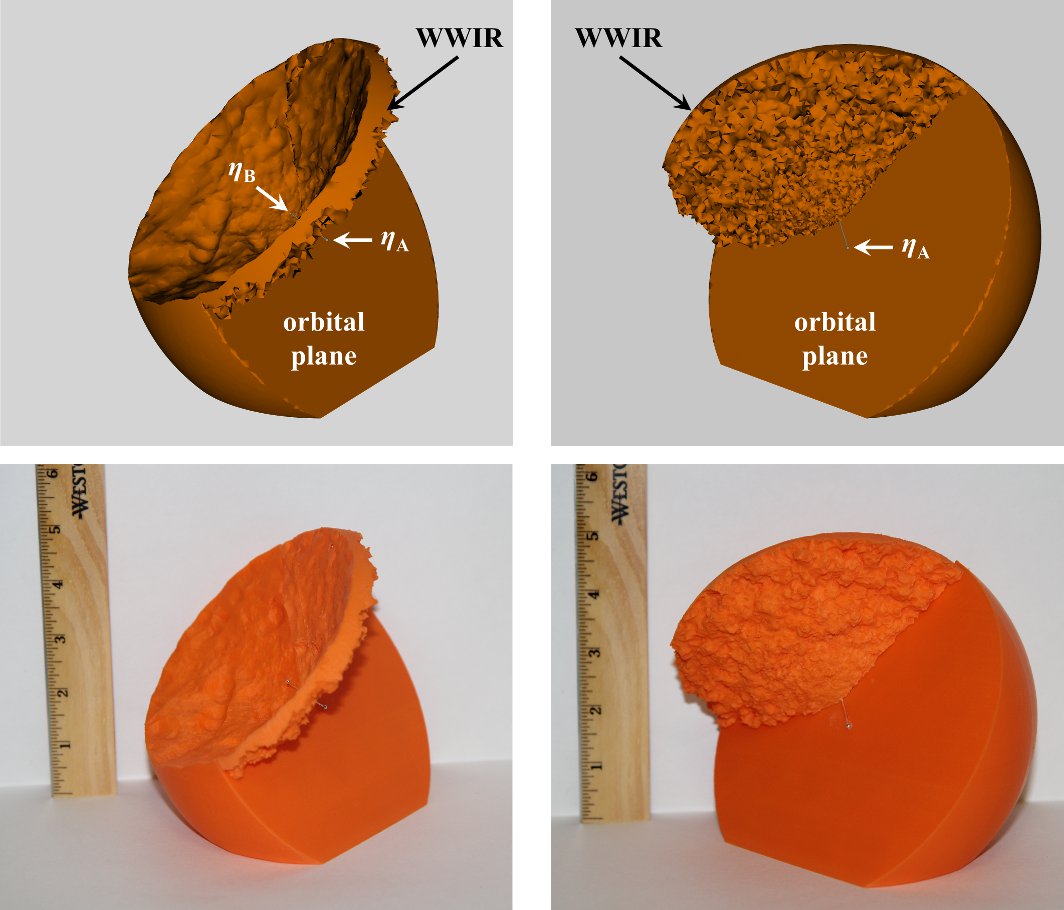}},
        3Dlights=CAD,
]{}{}{Fig8.u3d}
\caption{Comparison between the 3D rendering (top row) and 3D printed model (bottom row) of the Case~A SPH simulation at apastron. Columns present a (arbitrary) view of the WWIR with the observer looking into the \etaB wind cavity (left), and a view of the opposite side with the \etaB cavity opening away from the observer (right). Both views are looking down on the orbital plane. Click the image for a 3D interactive version of the model (Adobe Reader$^{\circledR}$ only). Pre-programmed views are available under the ``Views" menu in the 3D model toolbar. These include the projection of the system on the sky as viewed from Earth (view ``LOS", North up, East left), the primary wind and \etaB wind cavity only (view ``PrimaryWind"), and the wind-wind collision region plus stars only (view ``WWCR"). The physical diameter of the 3D printed models is approximately 6~inches (15.24~cm), as measured across the flat orbital plane through the centre of the model. At this scale, 1~inch (25.4~mm) corresponds to a distance $\approx 35.5$~au $\approx 5.31 \times 10^{9}$~km. The locations of the stars, WWIR, and orbital plane are indicated in the 2D figure. When in 3D interactive mode, right-click and select ``disable content'' to return to the 2D figure.}
\label{3dfig8}
\end{figure}

\begin{figure}
\centering \includemovie[
     3Dviews2=views_PeriastronHighMdot.tex,
        toolbar, 
        label=Fig9.u3d,
     text={\includegraphics[width=83mm]{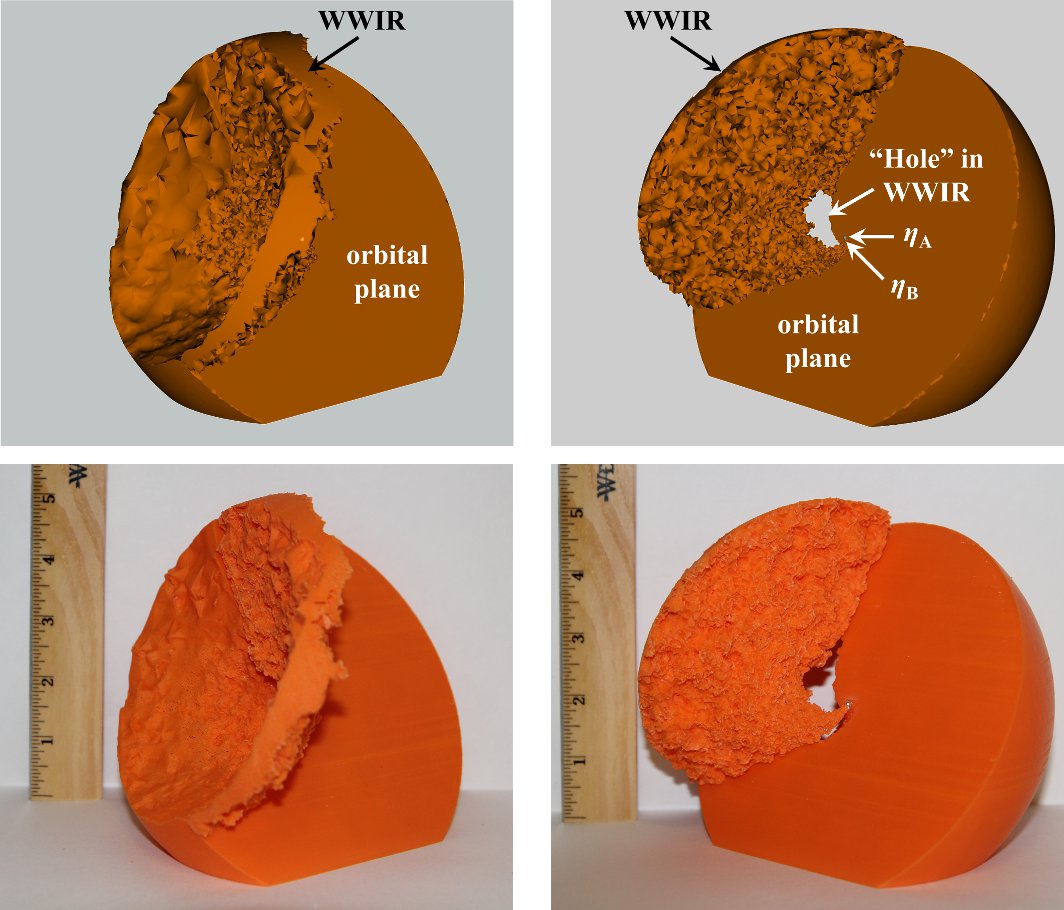}},
        3Dlights=CAD,
]{}{}{Fig9.u3d}
\caption{Same as Fig.~\ref{3dfig8}, but at periastron. Click image for a 3D interactive model (Adobe Reader$^{\circledR}$ only).}
\label{3dfig9}
\end{figure}

\begin{figure*}
\centering \includemovie[
     3Dviews2=views_Phase1p045HighMdot.tex,
        toolbar, 
        label=Fig10.u3d,
     text={\includegraphics[width=174mm]{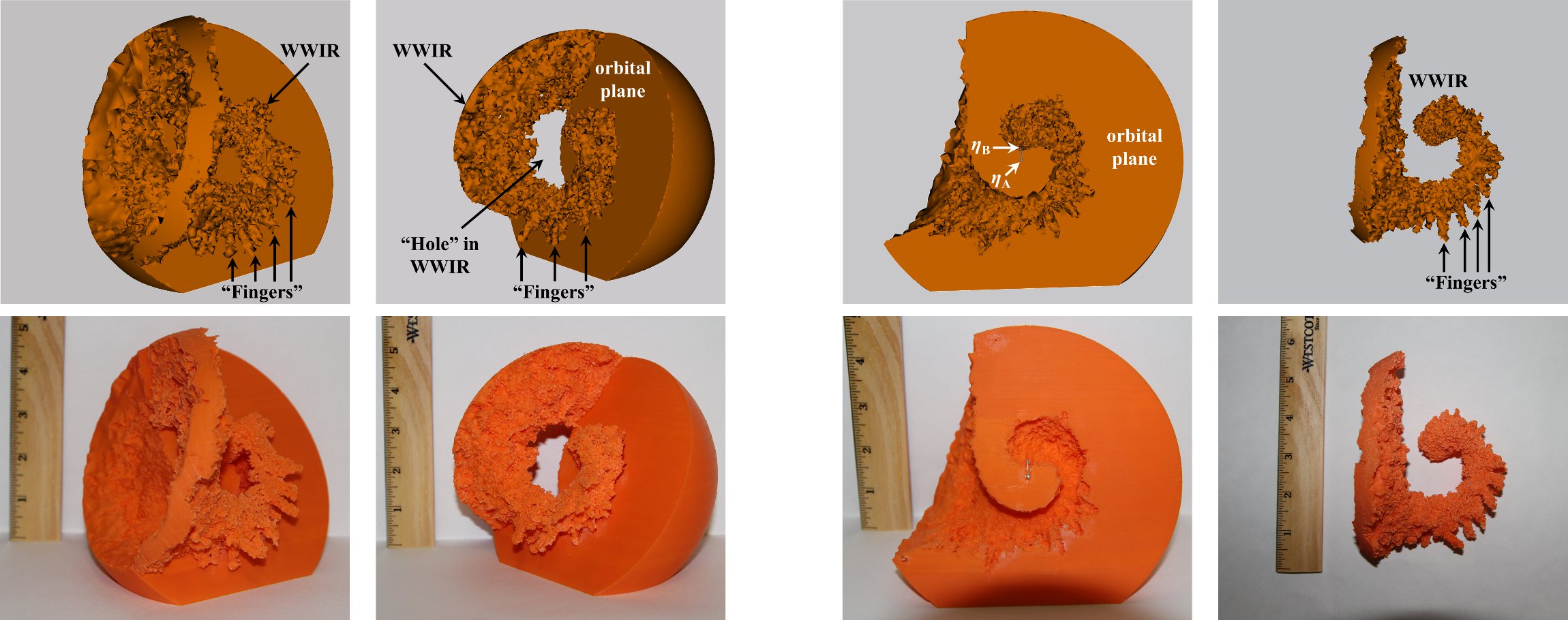}},
        3Dlights=CAD,
]{}{}{Fig10.u3d}
\caption{\emph{Left 4 panels}: Same as Fig.~\ref{3dfig8}, but at $\approx$~3~months after periastron ($\phi = 1.045$). \emph{Right 4 panels}: Comparison between a 3D rendering of the Case~A simulation (top row) and the 3D printed model (bottom row) for the two separable pieces that compose the model at $\phi = 1.045$. The left column shows the bottom half of the model with the dense \etaA wind and \etaB wind cavity. The right column shows the top half that consists solely of the WWIR. The locations of the stars, orbital plane, WWIR, WWIR hole, and WWIR fingers are indicated. Click the figure for a 3D interactive model (Adobe Reader$^{\circledR}$ only).}
\label{3dfig10}
\end{figure*}

\begin{figure}
\centering \includemovie[
     3Dviews2=views_ApastronLowMdot.tex,
        toolbar, 
        label=Fig11.u3d,
     text={\includegraphics[width=83mm]{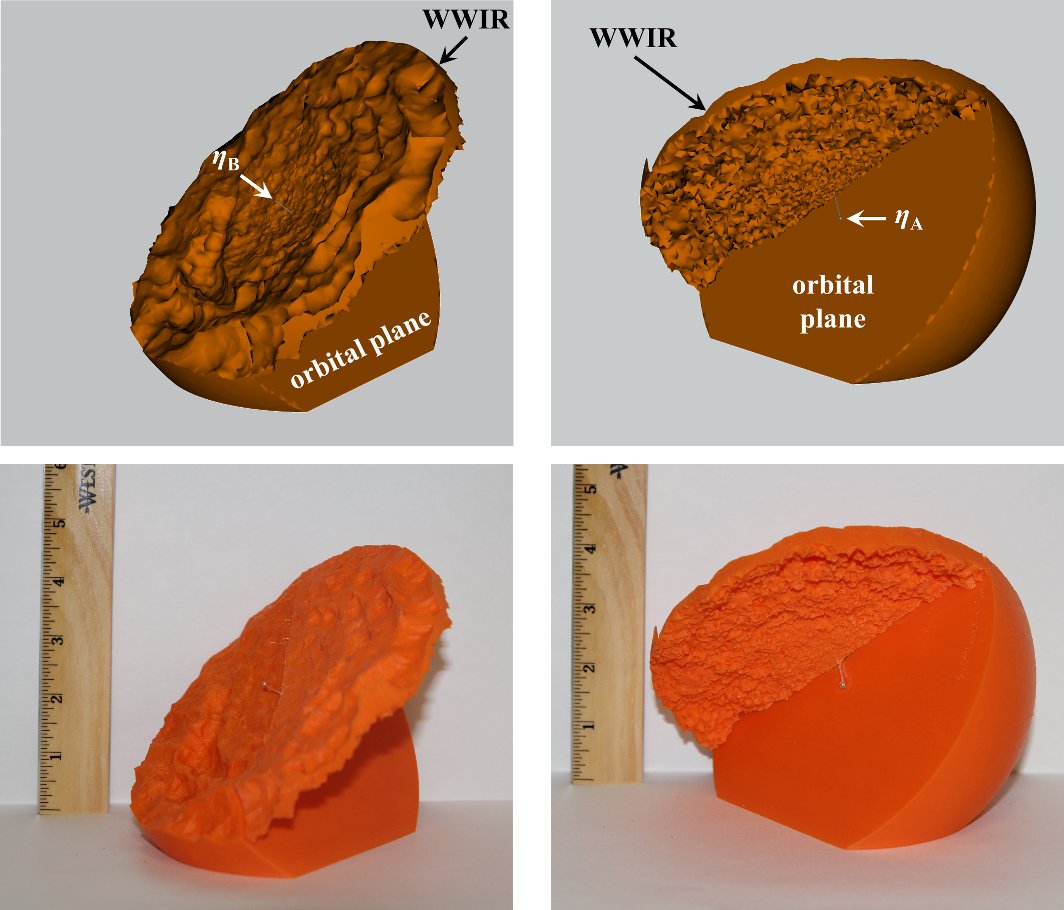}},
        3Dlights=CAD,
]{}{}{Fig11.u3d}
\caption{Same as Fig.~\ref{3dfig8}, but for Case~C. Click image for a 3D interactive model (Adobe Reader$^{\circledR}$ only).}
\label{3dfig11}
\end{figure}

\begin{figure}
\centering \includemovie[
     3Dviews2=views_PeriastronLowMdot.tex,
        toolbar, 
        label=Fig12.u3d,
     text={\includegraphics[width=83mm]{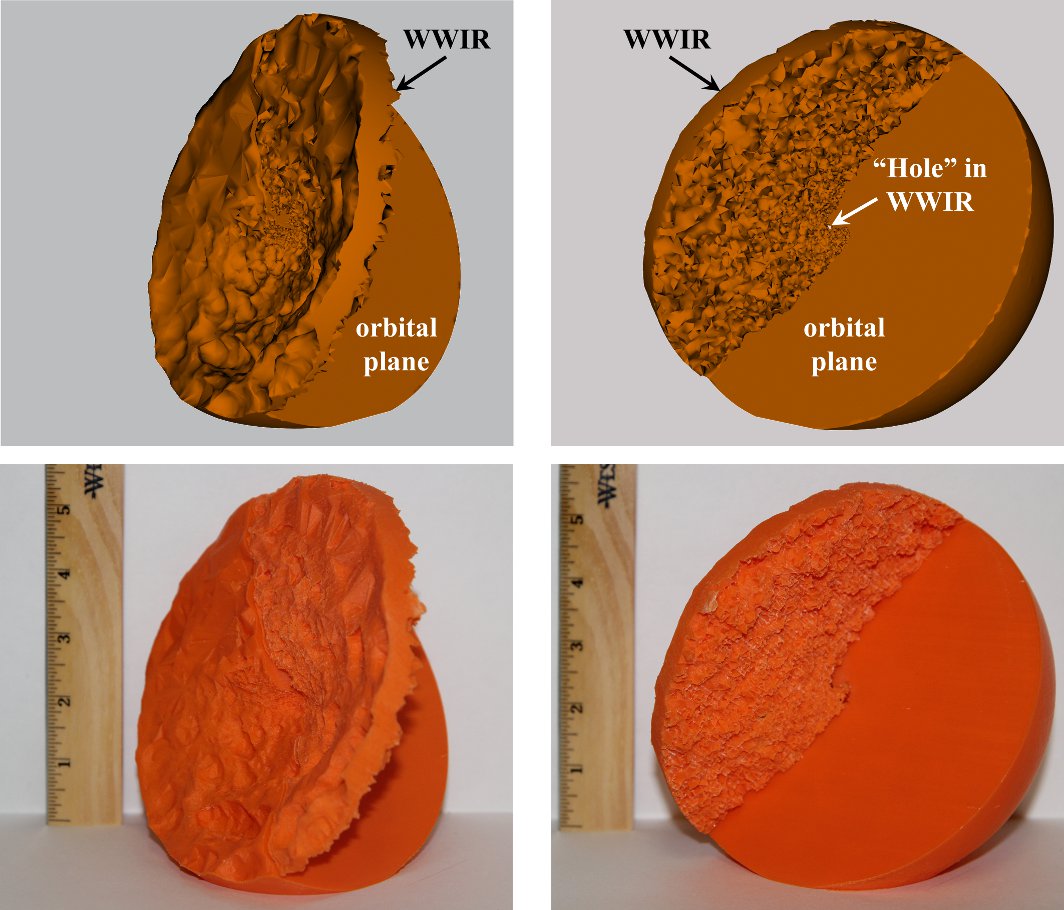}},
        3Dlights=CAD,
]{}{}{Fig12.u3d}
\caption{Same as Fig.~\ref{3dfig11}, but at periastron. Click image for a 3D interactive model (Adobe Reader$^{\circledR}$ only).}
\label{3dfig12}
\end{figure}

\begin{figure*}
\centering \includemovie[
     3Dviews2=views_Phase1p045LowMdot.tex,
        toolbar, 
        label=Fig13.u3d,
     text={\includegraphics[width=174mm]{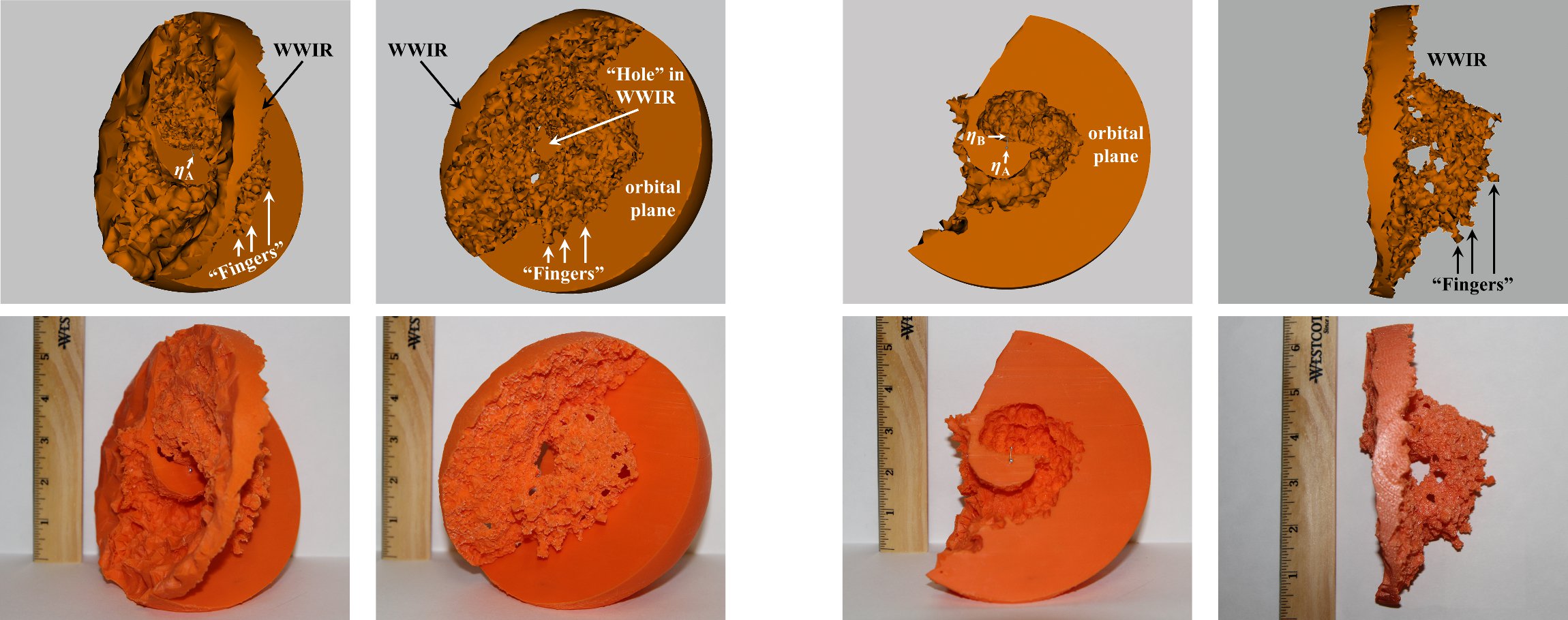}},
        3Dlights=CAD,
]{}{}{Fig13.u3d}
\caption{Same as Fig.~\ref{3dfig10}, but for Case C. Click the figure for a 3D interactive model (Adobe Reader$^{\circledR}$ only).}
\label{3dfig13}
\end{figure*}

\begin{figure*}
 \begin{center}
    \includegraphics[width=174mm]{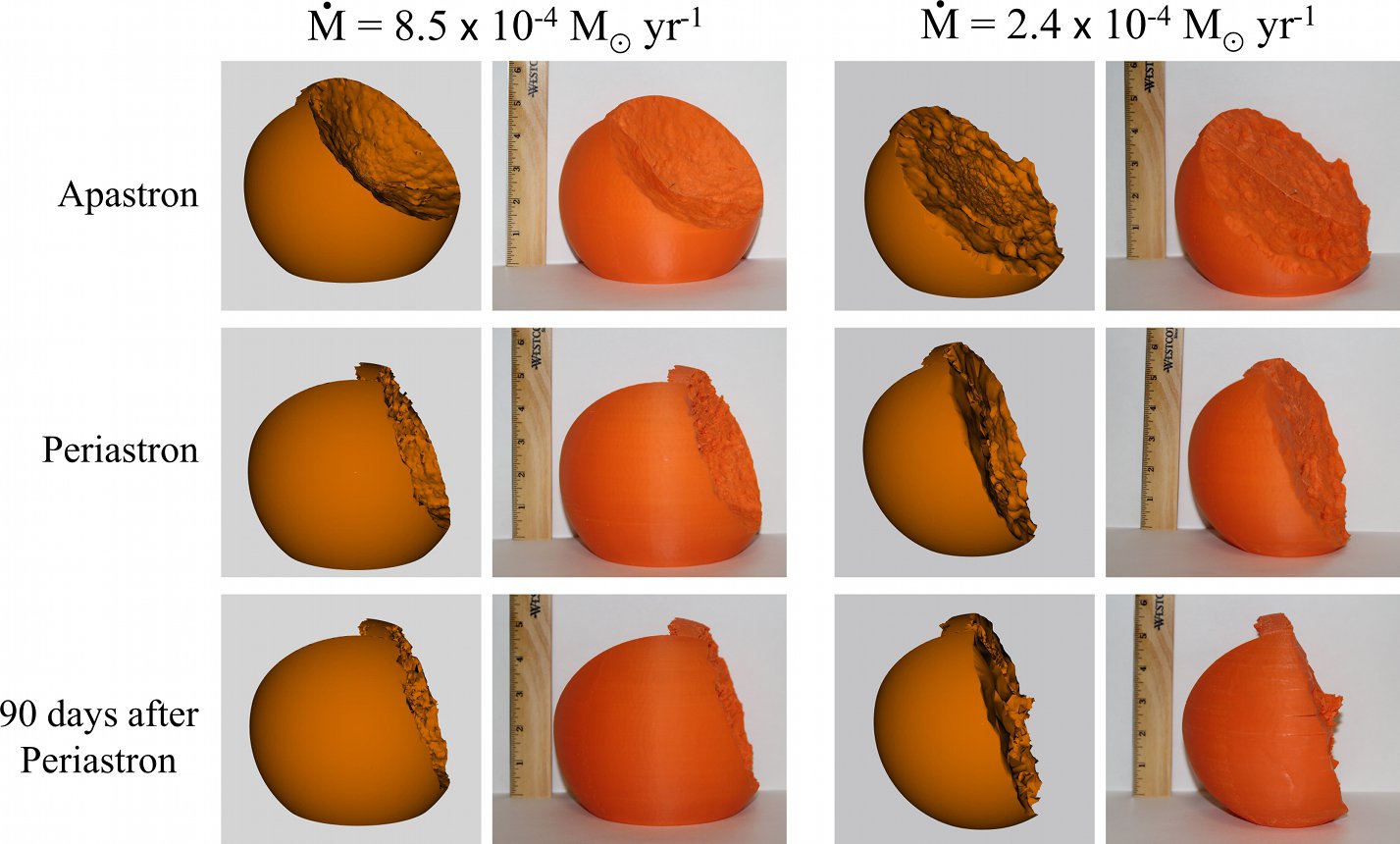}
    \caption{Comprehensive comparison between the 3D renderings and 3D printed models, with all models oriented as the system would appear on the sky to an observer on Earth (North is up, East is left). Rows (from top to bottom) present the three different orbital phases. The first two columns show, respectively, the 3D rendering and the 3D printed model, for Case~A.  The last two columns are the same as the first, but for Case~C.}\label{fig14}
 \end{center}
\end{figure*}

Figs.~\ref{3dfig8}--\ref{3dfig13} present our 3D printing results, showing a direct comparison between the actual 3D printed model and a 3D rendering of the model. In each 3D interactive model, the default starting view has the model oriented at the same position on the sky as the \ec binary \citep{madura12}, i.e. the starting view shows how an observer from Earth would see the system on the sky. Additionally, in the `Views' menu of the 3D graphics toolbar are options to display only the modified wind of \etaA, only the WWIR, or both together. For those unable to view the 3D interactive graphics, we include Fig.~\ref{fig14}, which shows how each model would appear on the sky to an observer on Earth. The STL files used for the prints are available as supplementary material in the online version of this article.

As demonstrated by the figures, the 3D printed models reproduce remarkably well all of the key features observed in the 3D interactive visualizations. We were pleasantly surprised that individual features in the 3D unstructured meshes, such as protrusions and trenches, were faithfully reproduced. One can make out in some locations the tetrahedral grid cells used in the mesh. Even the small protruding fingers that extend radially from the WWIR at $\phi =$1.045 were reproduced and remained mostly intact. However, we did have one delicate finger on the Case~A model accidentally break off, which was simply just glued back on (Fig.~\ref{3dfig10}).

The ability to hold and inspect the 3D printed models provides a new perspective on the WWIR's geometry and an improved sense of the scale of the different structures. One appreciates more just how large the WWIR is compared to the stars and stellar separation. The 3D models are also useful for constraining the observer's line-of-sight to the binary and help demonstrate why certain lines-of-sight are inconsistent with available observations. For example, as illustrated in Fig.~\ref{fig14}, at apastron, for the assumed orientation, our line-of-sight lies within the WWIR cavity, implying that any X-rays generated at the WWIR apex would be detectable to an observer at Earth. However, rotating the apastron model by $\sim 90^{\circ}$ or more places our line-of-sight through the dense, optically-thick primary wind, which would absorb any X-rays emitted from the WWIR apex. Thus, we may safely rule out such lines-of-sight.

The above is just one very simple example of how the 3D print models can be used. More importantly, we find that the 3D print models are extremely useful as a visual aid to help explain to non-\ec experts, and even non-astronomers, the 3D geometry and dynamics of the binary and WWIR. The 3D prints are a useful tool for illustrating concepts, relationships, and properties that are not easily conveyed by 2D, and even 3D, graphics.


\section{Discussion}\label{sec:Discussion}

The new 3D features discovered in our results may have some interesting implications for observational diagnostics of \ec and other highly-eccentric colliding wind binaries, such as WR~140. The hole near the apex of the WWIR that appears in the simulation snapshot at periastron will affect the generation of the shock-heated gas responsible for \ec's observed time-variable X-ray emission \citep{corcoran10,hamaguchi07,hamaguchi14}. Obviously, in areas where there is no wind-wind collision, there can be no shock-heated gas, and thus no thermal X-ray emission. Our 3D models imply that at periastron, there should be no thermal X-ray emission from along the trailing arm. Furthermore, as \MdotA is lowered, it should take longer for the hot gas, and thus X-ray emission, to vanish from the WWIR trailing arm as periastron is approached.

The WWIR hole at periastron also provides us with information about the amount and type of material in line-of-sight at that time. Since there is no longer a WWIR directly in line-of-sight at periastron, the column density of material between us and the stars is dominated by unshocked primary wind material that is flowing to fill the wind cavity carved by \etaB during the broad part of the orbit. Interestingly, this situation makes `wind-eclipses' by \etaA of various observed features (e.g. X-rays, \citealt{corcoran10}, and \HeII emission, \citealt{teodoro12}) easier to achieve at periastron for two reasons. First, since there is no longer a trailing arm to the WWIR, the overall size of the WWIR and volume of shock-heated gas at periastron is much smaller, and therefore easier to eclipse. Second, the lack of a trailing arm to the WWIR allows the dense \etaA wind to expand and fill the \etaB cavity in line-of-sight, increasing the size of the eclipsing wind photosphere.

The observational implications of the fingers that protrude from the WWIR at phases $\sim$3 months after periastron are unclear. Unfortunately, if real, they are too small to spatially resolve, even with \emph{HST}. Hot gas within the tubular fingers may produce X-rays, but the intensity of any such X-ray emission is likely to be small compared to that of X-rays generated near the WWIR apex. It is also unclear if the shocks responsible for producing the fingers could contribute to the \HeII $\lambda$4686 emission observed across \ec's periastron passage. Portions of the fingers located within \etaA's inner He$^{+}$ zone may be able to produce a small amount of \HeII emission if the shocks produce the required He$^{+}$-ionizing photons. However, this is very speculative without more detailed modeling.

It is difficult to constrain the observational implications of the WWIR fingers at this time because we lack a thorough understanding of their physical origins and properties. We currently speculate that the fingers arise as a result of instabilities at the interface between the two colliding wind shocks, which undergo rapid complex changes around periastron due to the high orbital eccentricity and changing wind directions. We point out that during the creation of the spiral WWIR, a typical wind-wind collision does not take place, due to the much faster orbital motion of the stars. The leading arm of the spiral WWIR is formed not by a head-on collision of two spherical winds, but rather by the fast receding \etaB wind colliding with the much denser and slower \emph{receding} \etaA wind. This collision still produces a pair of shocks, but the situation is now more analogous to that which occurs when a fast stellar wind interacts with a surrounding slower moving circumstellar shell \citep[e.g.][]{toala11, vanmarle12}, with the post-shock \etaA wind forming a thin, dense shell via radiative cooling, and the post-shock \etaB wind remaining hot and cooling adiabatically (see Fig.~\ref{fig7}).

Because of the high density contrast between the stellar winds, and because the faster \etaB wind is pushing into and accelerating the slower \etaA wind, different instabilities, including thin-shell \citep{vishniac83,vishniac94} and Rayleigh-Taylor \citep[RT][]{young01,toala11,vanmarle12}, are expected to arise and greatly distort/disrupt the spiral WWIR. Features very similar to the fingers that we observe in our models are visible in the wind-wind hydrodynamic simulations of \citet{toala11} (see e.g. their fig.~12). The protrusions that arise in the simulations of \citet{toala11} are a result of the thin-shell and RT instabilities. While not quite of the same magnitude as what we observe in our simulations, the 3D binary colliding wind simulations of \citet[][their fig.~10]{pittard09} and \citet[][their figs.~12 and 13]{parkin11} show somewhat similar features shortly after periastron, wherein the lower-density secondary wind penetrates into the denser primary wind due to thin-shell instabilities. Since \citet{pittard09} and \citet{parkin11} present mostly 2D slices of their simulations, we cannot be completely sure that the protrusions we observe are the same as the phenomenon shown in their figures. However, one reason such features may be stronger in our simulations is because our stellar winds have a larger density contrast than those in \citet{pittard09,parkin11}.

We find no obvious dependence of the fingers' properties on the resolution of the simulations in \citetalias{madura13}, with the exception that higher-resolution simulations appear to better resolve the instabilities in the WWIR, possibly leading to more fingers. Our interpretation that the fingers are due to strong instabilities that form in the WWIR around periastron makes qualitative sense, but we note that standard SPH schemes are notorious for under-resolving shocks and instabilities \citep{agertz07, price08}. Thus, our results should be interpreted with caution until a more detailed analysis is performed. Simulations using a grid-based method will help to determine if the fingers are an artifact of the SPH scheme. Such simulations will be the subject of future work. However, the fact that the hydrodynamic simulations of \citet{toala11,vanmarle12,pittard09}, and \citet{parkin11} were performed using grid-based methods and produced qualitatively similar phenomena supports the idea that the protrusions we observe are real. The timing and location of the fingers' appearance also makes physical sense on the grounds that they appear only during the rapid periastron passage, and occur only in the directions in which the fast wind of \etaB strongly collides with the slow, dense receding wind of \etaA.

\section{Summary and Conclusions}\label{sec:Summary}

In this work we present the first 3D prints of output from 3D SPH simulations of a dynamic astrophysical system. We demonstrate the methodology used to incorporate 3D interactive figures into a PDF document and the benefits of using 3D visualization and 3D printing as tools to analyze data from multidimensional numerical simulations. In this paper, we investigate the 3D structure of the WWIR in the the innermost regions of the \ec binary system. Below we summarize our most important results.

\begin{enumerate}[leftmargin=*, label=\arabic*.]

\item Several features in the overall geometry of the WWIR (e.g. changes in the opening angle due to different \MdotA, distortions of the cone cavity due to the orbital motion, and irregularities on the WWIR surface due to various instabilities) are fully and more easily appreciated with the help of 3D visualization.

\item The inability of \etaB's wind to collide with \etaA's downstream wind produces a large hole in the trailing arm near the WWIR apex at periastron. The size of this hole is directly connected with the wind momentum ratio, and therefore the WWIR opening angle. As expected, the higher \MdotA in Case~A causes a larger hole than that in Case~C. After periastron, the hole expands and propagates downstream along the vanishing WWIR trailing arm.

\item The faster orbital motion during periastron passage produces a much shallower spiral wind cavity on the back side of \etaA's wind compared to the nearly-axisymmetric conical cavity carved during the broad part of the orbit.

\item The 3D models present new `finger-like' features at phase 1.045. These protrusions extend radially, above and below the orbital plane, outward from the spiral WWIR for several au. The WWIR in Case~C exhibits large holes together with the protrusions. The presence of tubular cavities carved in \etaA's dense wind support the idea that these holes are actually indicative of protrusions with $\delta < 2$.

\item We speculate that the newly-identified finger-like protrusions are a result of thin-shell, RT, and other instabilities that arise where the receding fast \etaB wind collides with the dense, receding \etaA wind. Future simulations using grid-based methods are needed to confirm the existence of the fingers and determine their physical origin and properties.

\item The Case~C model shows that at phase 1.045, the outermost part of the leading arm of the `current' WWIR reaches and starts to merge with the remnant of the old trailing arm formed before periastron. In contrast, for Case~A, there is still a large gap between the `current' WWIR and the remnants of the trailing arm.

\item 3D models can be used to better visualize and constrain, for example, the observer's line-of-sight to the binary system, an important parameter to correctly interpret and model available observations. The 3D prints are also very useful for conveying complex ideas to non-experts.

\end{enumerate}

We demonstrate in this paper how software that generates 3D models, together with interactive 3D graphics in PDFs, can be used to produce publication-quality, scientifically instructive figures. Moreover, we show that 3D printed models reproduce extremely well the key features observed in the 3D interactive visualizations. Even if we have only touched on the possible applications of 3D printed models as a tool, this work helps highlight the important role 3D printing can play in understanding complex time-varying astrophysical systems. The ability to physically interact with the 3D models provides a completely new way to visualize, analyze, understand, and disseminate such 3D simulations. 3D print models are also extremely useful to show and explain to a non-expert or non-scientist the 3D geometry and dynamics of numerical simulations of astrophysical phenomena. Thus, 3D printing and visualization have the potential to improve the astrophysical community's ability to convey advances in our disciplines to the wider public, providing an opportunity for them to play a more active role in their learning by 3D printing various models. PDF is also the most widely-used, self-contained electronic document format, implying that funding agencies, governments, and the public can easily interact with instructive, 3D representations of our work \citep{barnes08}. We hope that this paper motivates others in the astrophysical community to pursue the use of 3D interactive visualization and 3D printing in their research and publications.


\section*{Acknowledgements}

TIM is supported by an appointment to the NASA Postdoctoral Program at the Goddard Space Flight Center, administered by Oak Ridge Associated Universities through a contract with NASA. We thank Fr\'{e}d\'{e}ric Vogt for very useful discussions on the incorporation of 3D interactive graphics into PDFs and 3D printing. We thank an anonymous referee for helpful comments.


\begin{flushleft}
\textbf{SUPPORTING INFORMATION}\\

Additional supporting information may be found in the online version of the article:\\

\textbf{3D Printable STL Files:} 3D print files for the models shown in Figs.~\ref{3dfig8}--\ref{fig14}.
\end{flushleft}

\bsp

\label{lastpage}

\end{document}